\documentclass[authoryear, preprint, 12pt]{elsarticle}

\usepackage{graphicx}
\usepackage{caption}
\usepackage{subcaption}
\usepackage{amssymb}
\usepackage{amsthm}
\usepackage{amsmath}
\usepackage{float}
\usepackage{bigfoot}
\usepackage{pdflscape}
\usepackage{adjustbox}
\usepackage{longtable}

\usepackage[utf8]{inputenc}

\usepackage{lineno}
\usepackage{lipsum}

\usepackage{natbib}

\usepackage{tabularx}
\newcolumntype{L}[1]{>{\raggedright\arraybackslash}p{#1}}
\newcolumntype{C}[1]{>{\centering\arraybackslash}p{#1}}
\newcolumntype{R}[1]{>{\raggedleft\arraybackslash}p{#1}}
\usepackage{multirow}

\usepackage[a4paper, margin=1in]{geometry}
\usepackage{hyperref}
\hypersetup{
	colorlinks   = true,    % Colours links instead of ugly boxes
	urlcolor     = blue,    % Colour for external hyperlinks
	linkcolor    = blue,    % Colour of internal links
	citecolor    = blue     % Colour of citations
}

\usepackage{setspace}
\usepackage{geometry}

\begin{document}

\begin{frontmatter}

\title{Revisiting the empirical fundamental relationship of traffic flow for highways using a causal econometric approach}

\author{Anupriya\corref{cor1}}
\author{Daniel J. Graham}
\author{Daniel H\"{o}rcher}
\author{Prateek Bansal}

\cortext[cor1]{Corresponding author.  Email address: anupriya15@imperial.ac.uk}
\address{Transport Strategy Centre, Department of Civil and Environmental Engineering \\  Imperial College London, UK
\vspace{-1cm}}

\begin{abstract}
The fundamental relationship of traffic flow is empirically estimated by fitting a regression curve to a cloud of observations of traffic variables. Such estimates, however, may suffer from the confounding/endogeneity bias due to omitted variables such as driving behaviour and weather. To this end, this paper adopts a causal approach to obtain the unbiased estimate of the fundamental flow-density relationship using traffic detector data. In particular, we apply a Bayesian non-parametric spline-based regression approach with instrumental variables to adjust for the aforementioned confounding bias. The proposed approach is benchmarked against standard curve-fitting methods in estimating the flow-density relationship for three highway bottlenecks in the United States. Our empirical results suggest that the saturated (or hypercongested) regime of the estimated flow-density relationship using correlational curve fitting methods may be severely biased, which in turn leads to biased estimates of important traffic control inputs such as capacity and capacity-drop. We emphasise that our causal approach is based on the physical laws of vehicle movement in a traffic stream as opposed to a demand-supply framework adopted in the economics literature. By doing so, we also aim to conciliate the engineering and economics approaches to this empirical problem. Our results, thus, have important implications both for traffic engineers and transport economists.\\
\end{abstract}


\end{frontmatter}

\newpage

\section{Introduction}
\label{S:1}

\begin{figure}[H]
  \centering
  \includegraphics[width= 0.8\linewidth]{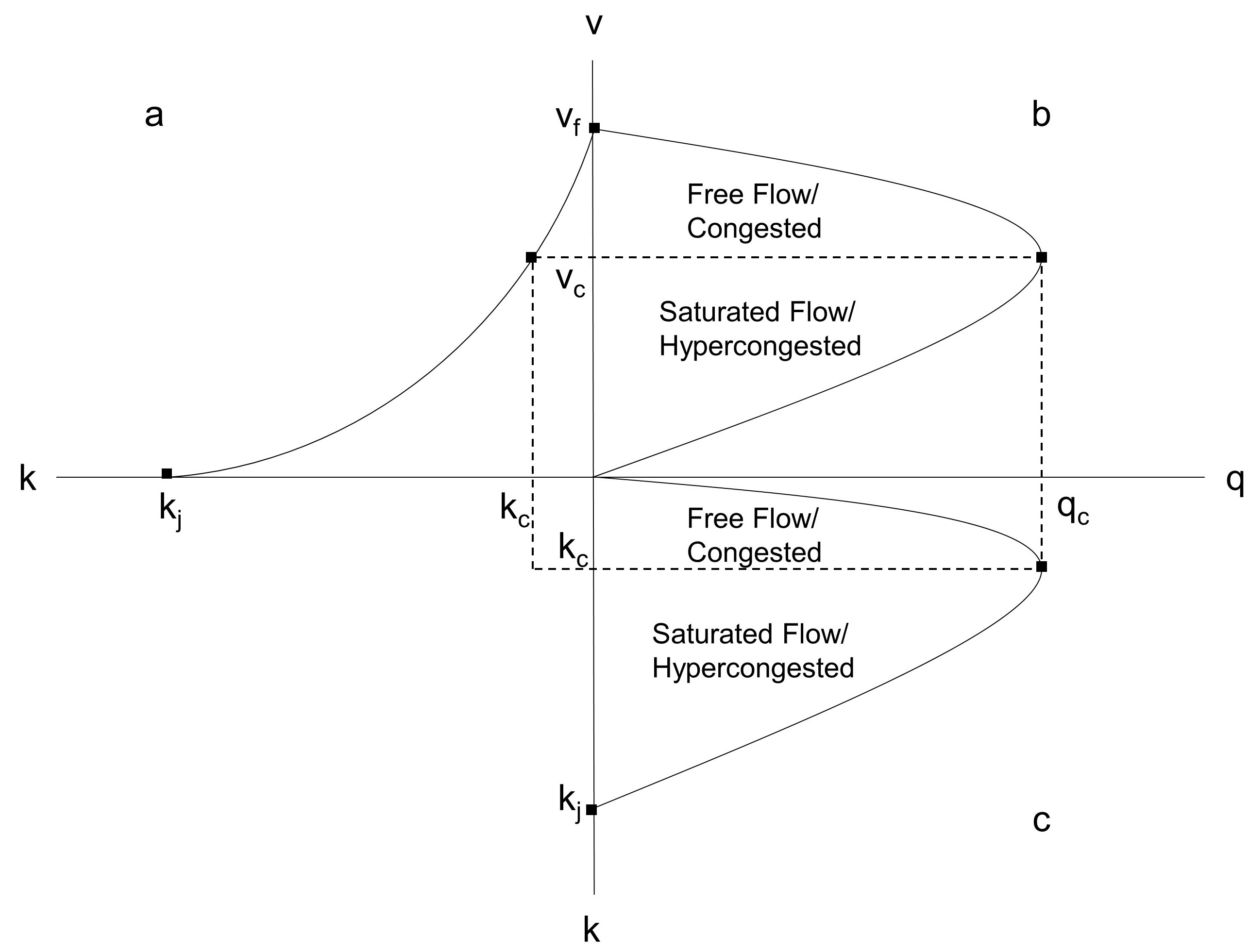}
  \caption{The fundamental diagram of traffic flow \citep[adapted from][]{Small2007}}\label{fig:FD}
\end{figure}

The standard engineering relationship between vehicular flow $q$, that is, the number of vehicles passing a given point per unit time, and density $k$, that is, the number of vehicles per unit distance in a highway section, as shown in quadrant (c) of Figure \ref{fig:FD}, is commonly known as the fundamental relationship of traffic flow. This relationship is defined based on the assumption that traffic conditions along the section are stationary, which means that the three key traffic variables, $q$, $k$ and average vehicular speed, $v$, are the same at each and every point in the highway section \citep{Daganzo1997,May1990}. Consequently, the relationship is basically estimated empirically by pooling observations from different cross-sections along the highway across different time-periods and fitting a regression curve to the point cloud. The estimation of such a curve follows from the engineers' interest in a general relationship to characterise the flow of traffic in a given facility. The fundamental relationship can be equivalently expressed as speed-density or flow-speed relationship, as shown in quadrants (a) and (b) of Figure \ref{fig:FD}, respectively.

Engineers assert that the estimated relationship is a property of the road section, the environment, and the population of travellers, because on an average, drivers show the same behaviour \citep{Daganzo1997}. We argue that this estimated relationship, however, is at best only \emph{associational and possibly spurious} due to several possible sources of endogeneity/confounding biases. For instance, there are many external observed and unobserved factors such as  driver behaviour, heterogeneous vehicles, weather and demand, that are correlated with the observed traffic variables \citep{Mahnke1999, Qu2017}, but are often omitted in the estimation of the fundamental relationship. Fitting a pooled ordinary least square regression curve to the observed scatter plot of traffic variables fails to adjust for the above-mentioned sources of confounding, which may bias the estimated relationship \citep{Wooldridge2010,Cameron2005}. The parametric limitations on functional form in regression further augments the bias in the estimated relationship.   

To address these shortcomings of the traditional approach, in this paper, we estimate the fundamental relationship between traffic flow and traffic density using a flexible causal statistical framework. In particular, we adopt a Bayesian non-parametric instrumental variables (NPIV) estimator \citep{Wisenfarth2014} that allows us to capture non-linearities in the relationship with a non-parametric specification without presuming the functional form and also adjust for any confounding bias via the use of instrumental variables (IVs). We validate this approach using traffic detector data from three highway bottlenecks located in California, USA\footnote{We choose highway bottlenecks over uniform highway sections to demonstrate that our approach not only delivers an unbiased fundamental relationship for a highway section but also correctly estimates capacity-drop, an important feature of the bottleneck section.}. Thus, the main contribution of this research lies in determining a novel causal (unbiased) relationship between traffic flow and traffic density for a highway bottleneck. 

To the best of our knowledge, this study presents the first application of \emph{causal inference} in empirical estimation of the fundamental relationship from an engineering perspective that is based on the physics of movement of vehicles. We emphasise that some economists have also adopted a causal framework for this problem in the past \citep[see, for instance,][]{Couture2018}. This framework is based on the interpretation of the speed-flow fundamental relationship as the supply curve for travel in a road section under stationary state traffic conditions \citep{Small2007,Walters1961}. We argue that in developing a causal understanding of the fundamental relationship, the economic representation of this model as a supply curve can lead to ambiguity. The economic interpretation seeks stationary state traffic conditions, which seldom exist, particularly under congested conditions.

Based on this type of economic representation, a recent empirical study by \cite{Anderson2020,Anderson2018} discards the existence of the hyper-congested\footnote{Whereas the engineering terminology for the backward bending region of the fundamental diagram is `saturated flow', economists call it `hypercongested' \citep{Small2003}.} part of the fundamental diagram (see Figure \ref{fig:FD} to locate the \textit{hypercongested} part). Note that for a highway bottleneck, there are two components of the hypercongested regime of the fundamental diagram: (i) the region representing capacity drop, that is, a sudden reduction in capacity of the bottleneck at the onset of upstream queuing, and, (ii) the region following the capacity drop where the flow-density or flow-speed relationship is backward bending as per the engineering literature. The absence of empirical evidence on the existence of both components of hypercongestion (that is, reduction in traffic flow with increasing traffic density or demand, \cite{Anderson2020,Anderson2018}) questions the relevance of hypercongestion as a motivating factor for traffic controls measures and congestion pricing strategies.\footnote{A few early studies in the engineering literature also report no capacity reductions \citep{Hall1990,Persaud1987,Newman1961}, however, their results have been found to be inconclusive owing to the methods adopted in these studies \citep{Cassidy1999a}.} 

Through our proposed causal framework, we also aim to conciliate this recently diverging strand from the economics literature with the well-established engineering knowledge on the existence of hypercongestion. Specifically, we contribute to the re-initiated debate on the existence of hypercongestion in highways and deliver novel causal estimates of capacity reduction in various highway bottlenecks. We emphasise that our estimates of the capacity reduction are derived from the estimated fundamental relationship itself, as opposed to the previous literature that uses different methodologies (e.g., change in cumulative vehicle count) to quantify the phenomenon \citep[see, for instance,][]{Cassidy1998,Oh2012,Srivastava2013}. Thus, our proposed approach provides a one-stop solution to estimate an unbiased fundamental relationship as well as its important features such as capacity and capacity-drop. As an important intermediate research outcome of this study, we also undertake a critical evaluation of the assumptions underlying the economists' treatment of the fundamental relationship as a supply curve for travel, which may lead to inconclusive empirical evidence on the existence of hypercongestion. 

The rest of this paper is organised as follows. Section \ref{S:2} reviews the relevant engineering literature, critically examining the theoretical foundations underlying the empirical estimation of the fundamental relationship between traffic variables. Additionally, we also review the literature on capacity-drop in highway bottlenecks. Section \ref{S:3} describes the chosen study sites and the corresponding traffic detector data and variables. Section \ref{S:4} details the model specification and the adopted methodology explaining how it addresses endogeneity bias in the context of the fundamental relationship. Section \ref{S:5} presents our results and benchmarks them against those derived using a standard non-parametric estimator without instrumental variables. Furthermore, we compare our findings with the relevant engineering and economics literature. Conclusions and implications are presented in the final section.

\section{Literature Review}
\label{S:2}

In this section, we discuss the theoretical foundations underlying the engineering approach to estimate the fundamental relationship and the key shortcomings of this approach. We also highlight how a causal econometric framework can be employed to obtain a more robust characterisation of the fundamental relationship.

\subsection{The empirical fundamental relationship}
\label{S:2.1}

\begin{figure}[t!]
  \centering
  \includegraphics[width= 0.7\linewidth]{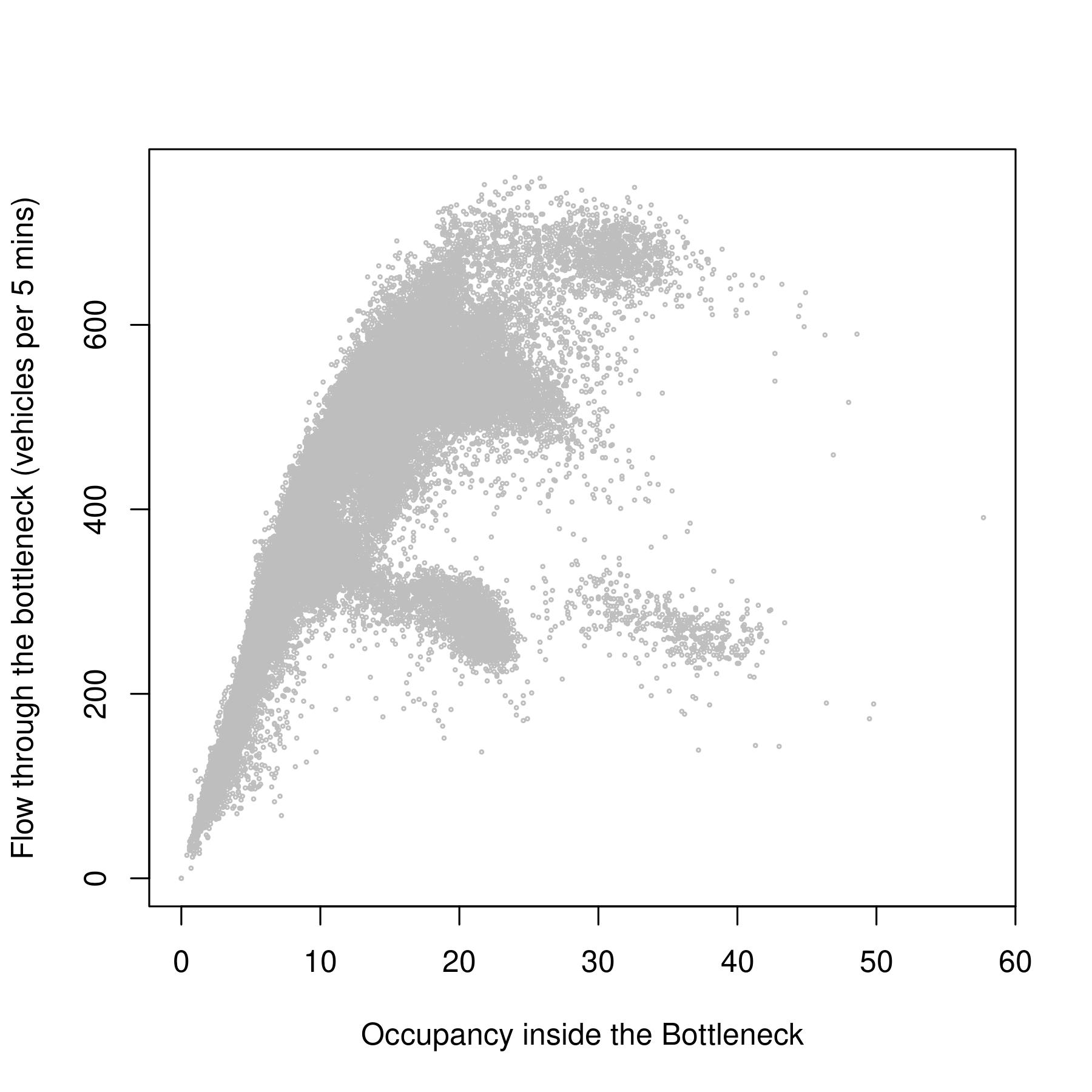}
  \caption{Conventional flow versus occupancy plot using detector data aggregated over 5 minutes.}\label{fig:SP}
\end{figure}

As discussed in the introduction, the fundamental relationship is empirically estimated using observations on traffic state variables that are averaged over time and/or space. The averaging of observations requires strict assumptions like \emph{stationary traffic} and \emph{homogeneous vehicles}. Note that empirical studies use occupancy, $o$, as a proxy for traffic density because traffic density cannot be measured directly \citep{Daganzo1997,May1990}\footnote{Occupancy is defined as the percentage of the sampling period for which vehicles occupy the detector}. For a contextual discussion, Figure \ref{fig:SP} shows a scattered plot of the measured flow versus occupancy from a traffic loop detector (aggregated over 5 minutes) located upstream of the Caldecott Tunnel in the SR24-W, California on several workdays. 

The conventional fundamental relationship is obtained by fitting a predefined curve to the point cloud of aggregated data. While numerous functional forms have been proposed \citep[see][for a review]{Hall1993}, most researchers agree that the flow-density relationship should either be triangular \citep{Newell1993,Hall1986} or parabolic \citep{Greenshields1935,HCM2016}. Proposed relations also include discontinuous models whereby the functions describing unsaturated and saturated traffic regimes do not come together to form a continuous curve \citep{Payne1977,Ceder1976,Drake1966}.  

As noted in previous studies, the data shows considerable scatter particularly in the saturated regime, which has led some researchers to question the existence of reproducible relationships in this regime \citep[refer to][for details]{May1990}. The challenges posed by this scatter in obtaining an accurate fundamental relationship has stimulated the growth of many interesting strands in the engineering literature: from understanding and accounting for the various sources of scatter \citep[see, for instance,][]{Cassidy1998} to introduction of stochastic fundamental relationships \citep{Wang2013} and fitting multi-regime models \citep[see, for instance,][]{Kidando2020}. The following subsections summarise the main findings from our review.

\subsubsection{Characterising the sources of scatter}
\label{S:2.1.1}

\cite{Cassidy1998} argued that data from brief periods of near-stationarity (that is, unsustained periods of time over which the traffic stream is marked by nearly constant average vehicle speeds) or non-stationary transitions do not always conform to near-stationary relations because such data points arise due to random fluctuations in the traffic variables. Further, \cite{Cassidy1998} showed that only data points from sustained periods of near stationary traffic condition conform to well-defined reproducible bivariate relations among traffic variables. \cite{Cassidy1998} and \cite{Daganzo1997} attribute such well-defined relationships to the same average behaviour of drivers when confronted with the same average traffic conditions. 

However, \cite{Coifman2014a} found that even under strict stationary state traffic conditions, aggregated measurements of flow, density (or occupancy) and speed may exhibit large scatter in the queued regime arising from: (i) erroneous measurements of flow due to non-integer number of vehicle headways in a given sampling period, (ii) averaging over a small number of vehicles during low flow periods, (iii) measurement errors due to detectors, (iv) the mixing of inhomogeneous vehicles (for instance, trucks and cars), and, (v) the mixing of inhomogeneous driver behaviours. Consequently, \cite{Coifman2014b} relaxed the requirement to seek out strict stationary state conditions by measuring the traffic state (flow and occupancy) for individual vehicles, followed by grouping of these measurements by similar lengths and speeds. For each group, \cite{Coifman2014b} derived the flow-occupancy relationship by connecting points corresponding to the median flow and median occupancy. \cite{Coifman2014b} argues that the use of median instead of the conventional use of mean controls for outliers arising from detector errors or uncommon driver behaviour.

We note that there are two main drawbacks of this approach. First, the method is highly data-intensive and requires microscopic-level measurements on individual vehicles. Second, although the method helps in estimating well-defined relationships for homogeneous vehicle classes, a traffic stream seldom consists of homogeneous vehicles only. This deficiency becomes a critical concern because the method does not clearly suggest how to obtain an aggregate relationship for a mix of vehicles from class-specific relationships, which is of general interest to devise traffic control measures and congestion pricing strategies.

\subsubsection{Stochastic fundamental relationships}
\label{S:2.1.2}

Consistent with \cite{Cassidy1998}, a series of other recent studies have attributed the observed scatter to random characteristics of traffic behaviour \citep{Qu2015,Chen2015,Mahnke1999,Qu2017,Muralidharan2011,Jabari2014,Sopasakis2004}. These studies suggest that the scatter arises due to various external factors such as heterogeneous vehicles, driver behaviour, weather conditions, and the random characteristics of demand. Previous empirical studies also demonstrate how failure to adjust for the stochastic characteristic of traffic flow variables in calibration of the fundamental relationship result into highly inaccurate models \citep{Ni2015,Wang2011,Li2012}. 

To account for these random characteristics, \cite{Wang2013} introduced a stochastic fundamental relationship in place of traditional deterministic models, in which they assume speed to be a random process of density and a random variable. Subsequently, \cite{Hadiuzzaman2018} and \cite{Kidando2019a} have used Adaptive Neuro Fuzzy Inference System and Markov Chain Monte Carlo (MCMC) simulations respectively to capture the uncertainty in parameter estimates of the fundamental relationship arising due from the stochastic behaviour of traffic.

We note that the baseline estimates of the flow-density or the speed-density curve derived within the stochastic framework are based on a pooled ordinary least square estimator. We argue that such estimates of the baseline curve may be confounded by the extraneous factors discussed in the literature. We explain this confounding bias in detail in Section \ref{S:4.2}. 

\subsubsection{Multi-regime models}
\label{S:2.1.3}

Traditional single-regime models \citep[for instance,][]{Greenberg1959,Pipes1966,Munjal1971} that assume a single, pre-defined shaped curve for the entire domain of the fundamental relationship have been found inaccurate because free-flow and congestion-flow regimes have different flow characteristics \citep{Ni2015,May1990,Hall1993}. As a consequence, multi-regime models have been introduced in the literature as a flexible alternative to increase the calibration accuracy of the fundamental relationship. Multi-regime models fit different regimes of the fundamental relationship with different pre-defined functional forms where regimes are separated by breakpoints or thresholds \citep{Edie1961,Drake1967,Sun2005}. Whereas two-regime models comprise of free-flow and congested-flow regimes, three-phase models additionally include a transitional regime between free-flow to congestion, which is consistent with Kerner's three-phase traffic theory \citep{Kerner2009}. 

In most studies, modellers pre-define the locations of breakpoints based on the subjective judgement, which may significantly affect the accuracy of the estimated multi-regime models \citep{Wang2011,Sun2005,Liu2019}. Studies such as \cite{Kockelman2001, Sun2005} and \cite{Kidando2019b} have thus focused on the estimation of these breakpoints based on a user-driven choice of the number of regimes as input. \cite{Kidando2020} proposed a fully data-driven Bayesian approach to estimate the breakpoints of multi-regime models, but did not account for the potential confounding biases. Our flexible non-parametric approach does not require any user inputs regarding the shape of fundamental relationship, automatically identifies such change-points in a data-driven manner and also accounts for the possible confounding biases.

\subsubsection{Research gaps and contributions}
\label{S:2.1.4}

Our review suggests that traffic engineers are interested in deriving a well-defined reproducible relationships between traffic variables that can be attributed to the same average behaviour of drivers under the same average traffic conditions. The developments in the engineering literature serve as an excellent starting point to understand the sources of variation in traffic state measurements that leads to large scatter, particularly in the congested regime of the fundamental diagram. While the previous studies on the stochastic fundamental diagram rightly argue that the scatter arises due to various external factors, they do not appropriately adjust for the confounding bias that may occur from these factors when estimating the aggregate (baseline) fundamental relationship. We argue that these factors are likely to be highly correlated with the observed traffic variables, thus, an ordinary least squares based estimation of the fundamental relationship may be biased (see Section \ref{S:4.2} for details). Moreover, most of the multi-regime models require user inputs for calibration and ignoring the endogeneity bias remains the concern. 

To fill these gaps in the literature, we introduce a methodological framework to estimate the fundamental relationship that can effectively control for confounding from the external sources identified in the literature, alongside adjusting for other inherent randomness in the data generating process, and produce a more general characterisation of traffic flow for a highway section under an average mix of traffic. The adopted fully flexible non-parametric specification for the fundamental relationship produces a multi-regime fundamental relationship without any prior assumptions on either the shape of the curve or the location of breakpoints. Moreover, as a by product of Bayesian estimation, we also quantify the uncertainty in the estimated relationship with credible intervals. Furthermore, important traffic control inputs for highway bottlenecks such as capacity and capacity drop are also obtained as a by-product of the estimation. In the rest of this section, we review the relevant literature on highway capacity and capacity drop to illustrate the importance of quantifying these parameters from the estimated fundamental diagram. 

\subsection{Highway Capacity and Capacity Drop}
\label{S:2.2}

Understanding the capacity of a highway section is critical in modelling of traffic flow in highways  \citep{Srivastava2013,Siebel2009,Laval2006}\footnote{Note that there are many potential definitions of capacity in the literature \citep[see][for a brief review]{Kondyli2017}. For instance, \cite{Cassidy2005} define capacity as the sustained flow discharged from all highway exits that are unblocked by spillover queues from downstream while the highway entrances are queued. \cite{Oh2012} define capacity as the maximum discharge flow of vehicles that persist for 5 minutes in a free-flow state.}, particularly those with bottlenecks (such as lane drops and merges, among others). This is because, the highway capacity at the bottleneck location may be insufficient for traffic demand during peak hours and hence, traffic jams may occur. Capacity drop is thus defined as the drop in discharge flow through a bottleneck, when it is activated with an increase in demand. The activation of a bottleneck is marked by onset of queuing upstream of the bottleneck \cite{Yuan2015,Oh2012,Chung2007,Cassidy1999a}. The literature also acknowledges capacity drop as a \textit{two-capacity} phenomenon of active bottlenecks and relate it to the discontinuity observed at capacity flows near saturation point in the flow-density or flow-speed fundamental diagram. Several empirical observations of capacity drop ranging between 2 percent -25 percent are found in the literature. Table \ref{tab:LR} presents a summary of the capacity drop reported in the literature for different highway sections with varying bottleneck type. Based on the behavioural theory of traffic flow, \cite{Daganzo2002} attributes the capacity drop to a loss of \emph{motivation} among drivers, that is, these drivers presumably loose their willingness to drive at high speeds with small headways. In addition, \cite{Laval2006} and \cite{Leclercq2011} suggest that capacity drop occurs due to voids in the traffic caused by lane changing. Past researchers also relate the differences in capacity drop values with the number of lanes, severity of stop-and-go waves and speeds in congestion \cite{Oh2012,Oh2015,Yuan2015}.

\begin{table}[H]
\caption{Summary of key literature on the existence capacity drop in highways.}
\label{tab:LR}
\centering
\begin{adjustbox}{width= 1\textwidth}
\begin{tabular}{lllr}
\hline
Study & Location & Type & Capacity Drop (\%) \\
\hline
\cite{Banks1990} & I-8, San Diego & on-ramp merge & -0.42 to 1.11\\ 
\cite{Hall1991} & Queen Elizabeth Way, Toronto & on-ramp merge & -7.76 to 10.36\\ 
\cite{Banks1991} & Multiple Sites, San Diego & merge/ lane drop/ weave & 1.8 to 15.4\\ 
\cite{Persaud1998} & Multiple Sites, Toronto & on-ramp merge & 10.6-15.3\\ 
\cite{Cassidy1999a} & Multiple Sites, Toronto & on-ramp merge & 4 to 10\\ 
\cite{Bertini2004} & US-169, Minneapolis & on-ramp merge & 2 to 5\\ 
\cite{Bertini2005} & M4, London & merge & 6.7 to 10.7 \\
\cite{Cassidy2005} & I-805, San Diego & on-ramp merge & 8.3 to 17.3 \\ 
\cite{Chung2007} & I-805, San Diego & on-ramp merge & 5 to 18\\ 
 & SR-24, California & on-ramp merge, lane reduction & 5.1 to 8.5\\ 
 & Gardiner Expressway, Toronto & on-ramp merge, horizontal curve & 3 to 12\\ 
\cite{Leclercq2011} & M6, UK & merge & 25\\ 
\cite{Oh2012} & Multiple Sites, California & on-ramp merge & 8 to 16.5\\
\cite{Srivastava2013} & US-169, Minneapolis & on-ramp merge & 8 to 15\\ 
\cite{Jin2015} & I-405, California & merge & 10.5\\ 
\cite{Anderson2020} & Multiple Sites, California & lane reduction & 0\\ 
\hline
\multicolumn{4}{l}{\footnotesize{*This table has been adapted from \cite{Oh2012}.}}
\end{tabular}
\end{adjustbox}
\end{table}

The existence of capacity-drop in a highway section is well-recognised in the transportation literature and has been a long-standing rationale for application of traffic controls, such as, ramp metering \citep{Cassidy2005,Smaragdis2004,Diakaki2000}, and highway pricing and tolls \citep{Hall2018a,Hall2018b,Small2013,Newbery1989,Boardman1977,Walters1961} to regulate demand. We note that the methods adopted in the literature to quantify capacity drop differ substantially from each other. For instance, \cite{Banks1990}, \cite{Persaud1998}, \cite{Zhang2004} and many more study minute-to-minute variability in traffic flows to infer decrease in high traffic flow levels. Studies like \cite{Bertini2005} and \cite{Cassidy2005} use cumulative vehicle counts to infer the reduction in flow at downstream detectors relative to upstream detectors. \cite{Srivastava2013} study changes in bottleneck flows with respect to upstream density. 

However, the capacity-drop phenomenon has recently been called into question in the urban economics literature. \cite{Anderson2020,Anderson2018} study the changes in capacity of a highway section with a bottleneck during periods of high demand for three bottlenecks in California and conclude that there is no evidence of capacity drop, hence, hypercongestion, in the absence of any infrastructure-related shocks (for instance, lane closures, traffic incidents and so on) and weather-based shocks. Consequently, they question the relevance of hypercongestion in the design of traffic controls and congestion pricing. 

The capacity drop estimates in engineering studies are based on only a certain few days of observations. \cite{Anderson2020} instead use data from several hundred days and adopt an event-study design to measure changes in highway capacity before and after of queue formation averaged over all days. It is important to note that \cite{Anderson2020} select a speed threshold of 30 mph at the upstream detector closest to the bottleneck to detect the onset of upstream queuing. 

We argue that inferring the actual moment of queue formation using a speed threshold for upstream detectors may lead to ambiguity. We instead reevaluate the capacity drop phenomenon by deriving estimates of capacity drop from the causal estimate of the flow-occupancy relationship at the bottleneck location. However, as rightly suggested by \cite{Anderson2020}, we use several months of observations to separate capacity drop from minute-to-minute fluctuations in traffic flow.

\section{Data and Relevant Variables}
\label{S:3}

We make use of traffic data from three standard highway sections with distinct geometry, each having a single and clearly identified active bottleneck located at its downstream end. At all of the chosen sites, slowing down of traffic and queuing is observed at the bottleneck location. The associated high-quality data is collected via a series of loop detectors installed at various locations along the highway, which measure traffic flow and occupancy averaged over every 5-minute duration. The data is maintained by the California Department of Transportation (Caltrans) and made publicly available through their Performance Measurement System (PeMS) website\footnote{Performance Measurement System (PeMS) website: \url{http://pems.dot.ca.gov/}}. 

\subsection{Study Sites}
\label{S:3.1}

\subsubsection{Site 1}

The first bottleneck that we study is located in the westbound direction of the California State Route 24 (SR-24) at the Caldecott Tunnel in Oakland, California. The SR-24 connects suburban Contra Costa County in the East Bay region of the San Francisco Bay Area with the cities of Oakland and San Francisco in the west. During the period 2005-2010, the Caldecott tunnel was operated  with two reversible lanes carrying the westbound traffic in the morning and the eastbound traffic in the afternoon and evening. Thus, for afternoon and evening hours of the above period, the location features an active bottleneck in the westbound direction with the number of lanes decreasing from \textit{four to two}. As the traffic approaches the tunnel, traffic delays being quite common at this location \citep[previously studied by][among others]{Chin1991,Chung2007,Anderson2020}. We use observations on the westbound traffic in the time period 12:00-24:00 hours on weekdays in the months of June-August during 2005-2010. This highway section is well-isolated, that is, located well away from any major downstream intersection. Consequently, we assume that this section allows us to study traffic dynamics arising solely from the presence of the bottleneck, without being affected by any downstream influences.

\subsubsection{Site 2}

Our second study site is located in the eastbound direction of the California State Route 91 (SR-91). SR-91 connects several regions of the Greater Los Angeles urban area in the west with the Orange and Riverside Counties in the east. At the location where two-lane traffic from the Central Avenue- Magnolia Centre in the Riverside Country merges with its three-lane eastbound traffic, it features an active merge bottleneck \citep[previously studied by][]{Oh2012}. This bottleneck appears as one of the top 100 bottlenecks in California enlisted on the PEMS website with queuing and delays being quite common at this location during morning and evening hours. We use observations on the eastbound traffic in the time period 06:00-12:00 hours on weekdays in the months of June-August during 2009-2014. This highway section is also well-isolated, thus, the traffic dynamics arising within the section arise solely from the presence of the bottleneck.

\subsubsection{Site 3}

The third bottleneck that we study is located in the eastbound direction of California State Route 12 (SR-12). SR-12 connects the Sonoma, Napa, and Solano Counties, following which it merges with Interstate 80 (I-80) which continues north towards Sacramento. At a location just west of I-80, the number of lanes in the highway drops from two lanes to one. This site has been previously studied by \cite{Anderson2020} who suggest that the lane-drop results in the formation of an active bottleneck with queues that are often very long. We use observations on the eastbound traffic in the time period 12:00-24:00 hours on weekdays in the months of May-August during 2018-2019. 

Further downstream of the lane drop, the SR-12 merges with I-80. However, \cite{Anderson2020} note that the dynamics within this section are not affected by this merge. We adopt this section for further analysis as it offers an interesting avenue to verify the effect of downstream influences on the estimated fundamental relationship within the bottleneck.

\subsection{Relevant Variables}
\label{S:3.2}

\begin{figure}[tp]	
	\centering
	\begin{subfigure}[h]{1\linewidth}
	\centering\includegraphics[width=\linewidth]{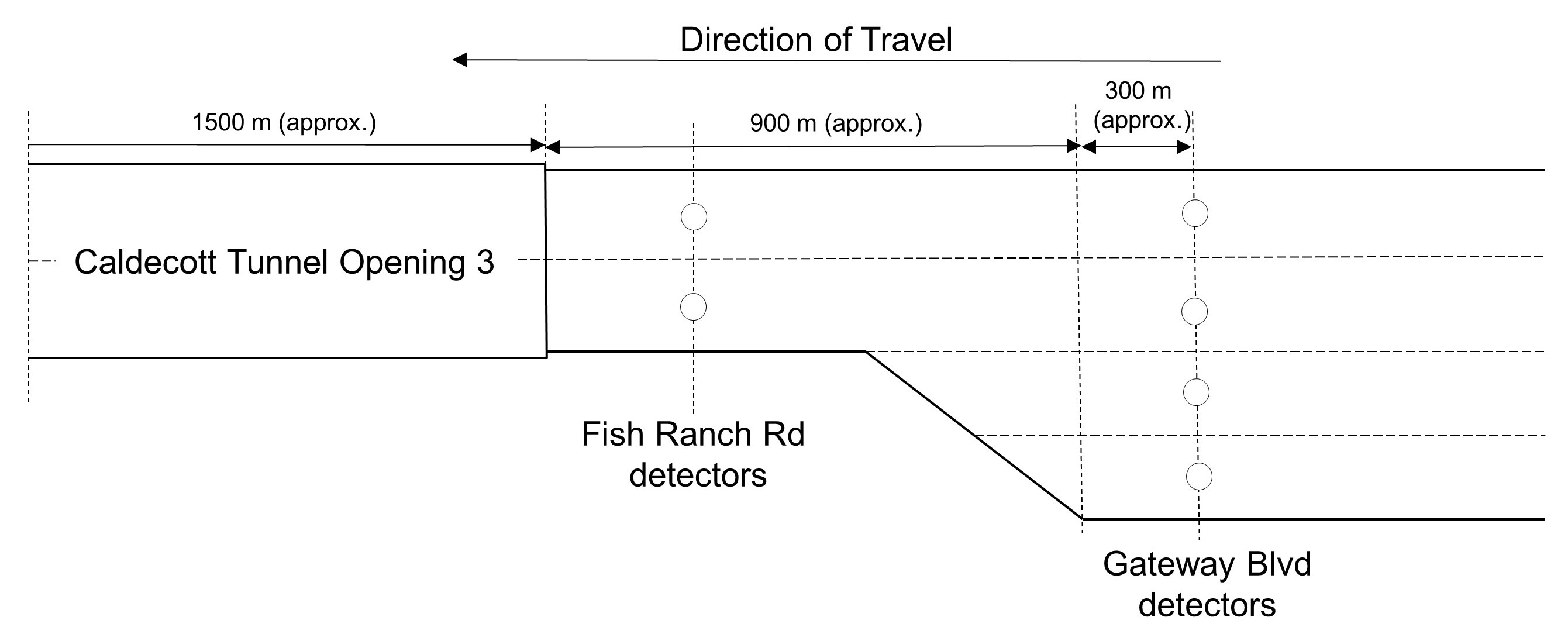}
	\caption{A lane-drop bottleneck in the westbound SR-24 in Oakland, California.}
	\label{fig:M(a)}		
	\end{subfigure}
	\hfill
	\begin{subfigure}[h]{0.8\linewidth}
	\centering\includegraphics[width=\linewidth]{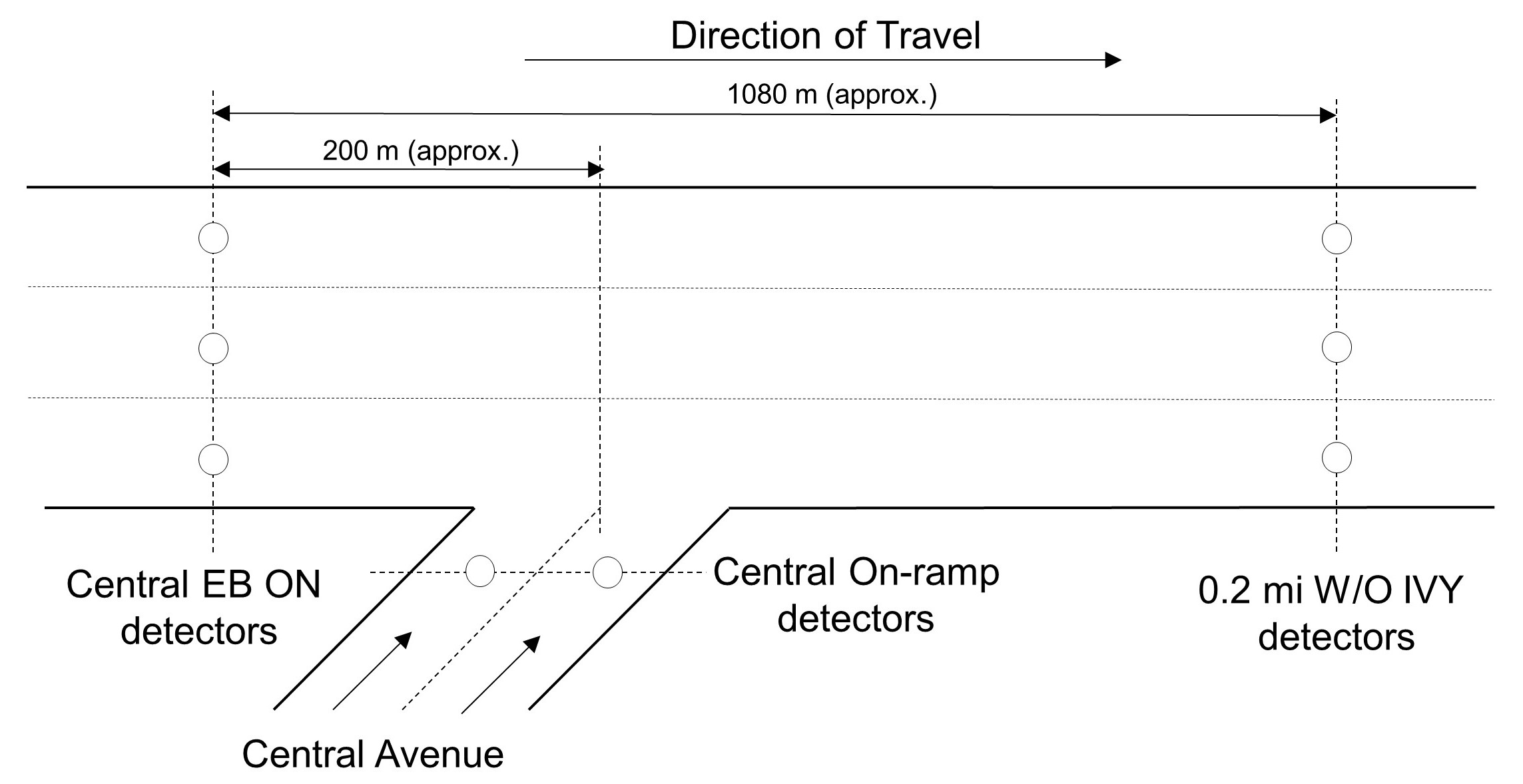}
	\caption{A merge bottleneck in the eastbound SR-91 in Riverside, California.}
	\label{fig:M(b)}
	\end{subfigure}
	\hfill
	\begin{subfigure}[h]{0.6\linewidth}
	\centering\includegraphics[width=\linewidth]{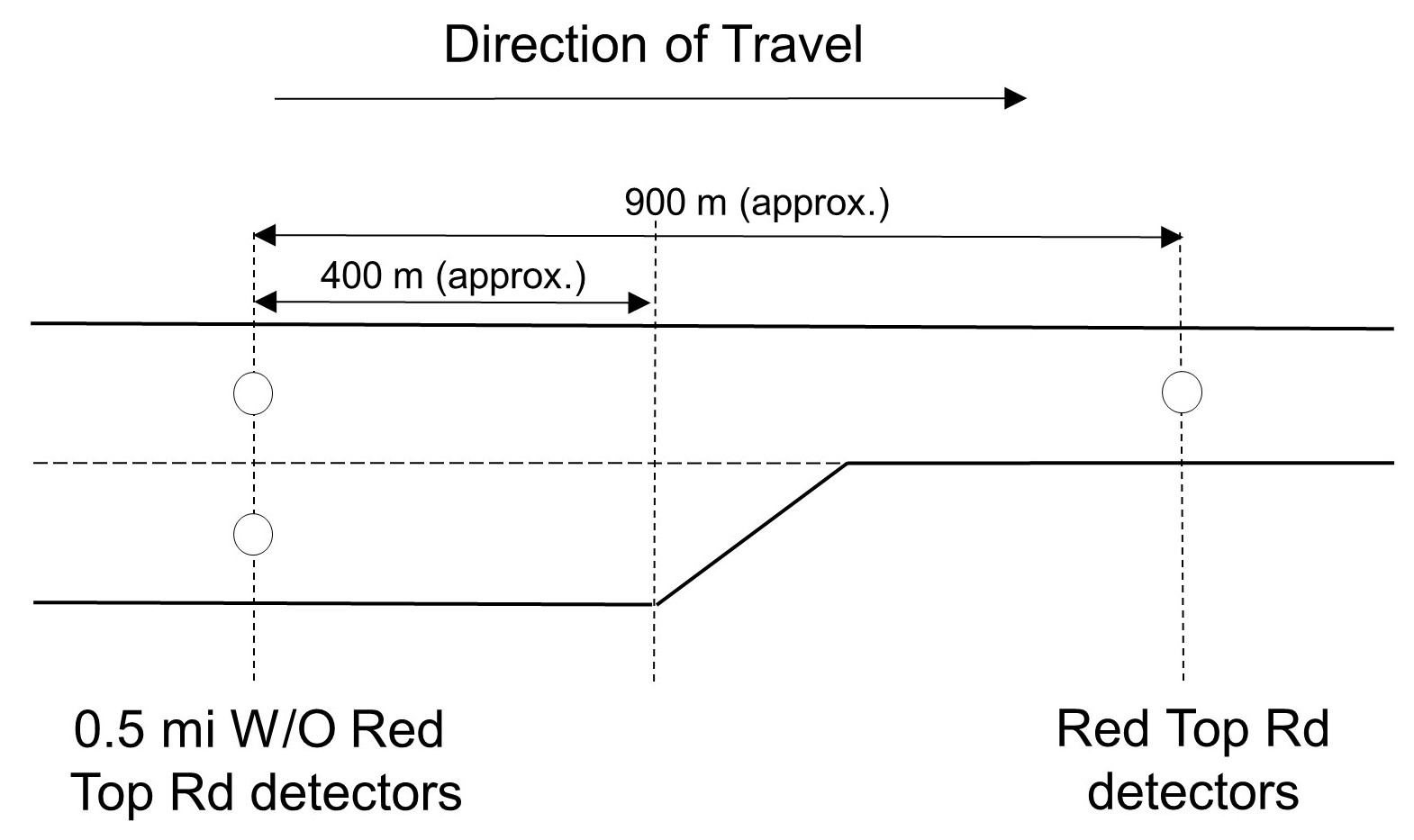}
	\caption{A lane-drop bottleneck in the eastbound SR-12 in Solano, California.}
	\label{fig:M(c)}
	\end{subfigure}
\caption{Schematic representation of the study sites.} \label{fig:Site}
\end{figure}

A schematic representation of the three bottlenecks, along with the location of detectors that we use to obtain the relevant data, is shown in Figure \ref{fig:Site}. 

For the first site, we observe a set of two detectors downstream of the lane-drop (that is, within the bottleneck) and four detectors upstream to it. For the second site, we observe a set of three detectors downstream of the merge (that is, within the bottleneck) and three detectors upstream to it. For the third site, we observe one detector downstream of the lane-drop (that is, within the bottleneck) and two detectors upstream to it. 

It is also worth emphasising that for all the three bottlenecks, there are no reasonable alternative routes to the highway section for the analysed traffic. We can thus assume that, on an average, driver population using the section during the study period does not differ substantially on weekdays. Table \ref{tab:sumstats} summarises the relevant variables from the three sites that are used in this study. 

\begin{table}[H]
\caption{Summary statistics for variables used in this analysis.} \label{tab:sumstats}
\centering
\begin{subtable}[h]{0.8\textwidth}
\caption{Site 1: SR-24 westbound.}
\centering
\adjustbox{max width=\textwidth}{
\begin{tabular}{llrrrrr}
  \hline
  Variable & Detectors & Obs. & Min & Max & Mean & Std.Dev \\ 
  \hline
  Traffic Flow (veh/5mins) & Gateway Blvd &	54432 & 0.00 & 595.00 & 249.58 & 77.57 \\
  Occupancy & Gateway Blvd & 54432 & 0.00 & 73.60 & 18.82 & 18.16 \\
  Traffic Flow (veh/5mins) & Fish Ranch Rd & 54432 & 0.00 & 760.00 & 383.95 & 149.57 \\
  Occupancy & Fish Ranch Rd & 54432 & 0.00 & 57.70 & 12.29 & 7.19 \\
  \hline 
\end{tabular}}
\bigskip
\end{subtable}
\hfill
\begin{subtable}[h]{0.8\textwidth}
\caption{Site 2: SR-91 eastbound.}
\centering
\adjustbox{max width=\textwidth}{
\begin{tabular}{llrrrrr}
  \hline
  Variable & Detectors & Obs. & Min & Max & Mean & Std.Dev \\ 
  \hline
  Traffic Flow (veh/5mins) & Central EB ON & 27936 & 102.00 & 593.00 & 372.21 & 53.03 \\
  Occupancy & Central EB ON & 27936 & 3.60 & 57.90 & 11.20 & 5.08 \\
  Traffic Flow (veh/5mins) & W/O IVY & 27936 & 152.00 & 628.00 & 393.55 & 69.77 \\
  Occupancy & W/O IVY & 27936 & 3.60 & 63.20 & 10.89 & 3.09 \\
  Traffic Flow (veh/5mins) & Central On-ramp & 27936 & 0.00 & 131.00 & 35.99 & 33.33 \\
  \hline
\end{tabular}}
\bigskip
\end{subtable}
\hfill
\begin{subtable}[h]{0.8\textwidth}
\caption{Site 1: SR-12 eastbound.}
\centering
\adjustbox{max width=\textwidth}{
\begin{tabular}{llrrrrr}
  \hline
  Variable & Detectors & Obs. & Min & Max & Mean & Std.Dev \\ 
  \hline
  Traffic Flow (veh/5mins) & W/O Red Top Rd & 24908 & 0.00 & 210.00 & 102.13 & 43.08 \\
  Occupancy & W/O Red Top Rd & 24908 & 0.00 & 74.90 & 18.64 & 19.44 \\
  Traffic Flow (veh/5mins) & Red Top Rd & 24908 & 0.00 & 173.00 & 99.68 & 43.71 \\
  Occupancy & Red Top Rd & 24908 & 0.00 & 69.10 & 11.41 & 6.84 \\
  \hline
\end{tabular}}
\bigskip
\end{subtable}
\end{table}

\section{Methodology}
\label{S:4}

This section is divided into four subsections. In the first subsection, we discuss the model specification. In the second subsection, we explain potential endogeneity bias in estimation of the fundamental relationship. In the penultimate subsection, we briefly review NPIV methods in the literature and describe the Bayesian NPIV method in the context of this study. In the concluding section, we benchmark the performance of the Bayes NPIV estimator against state-of-the-art estimators in a Monte Carlo study and illustrate its superiority in adjusting for endogeneity bias and recovering complex functional forms.

\subsection{Model Specification}
\label{S:4.1}

We estimate a causal relationship between occupancy inside the bottleneck, $o^b_{it}$, in the five-minutes interval $i$,  $i=1,..,N$, on a particular day $t$, $t=1,...,T$, and the flow through the bottleneck, $q^b_{it}$. We consider $q^b_{it}$ to be a function of $o^b_{it}$, conditional on the properties of the infrastructure, the environmental conditions and the average behaviour of drivers and vehicles. 

\begin{equation}
\label{eq:spec}
q^b_{it} = S^{b}(o^b_{it}) + \delta^{b}_{it} + \xi^{b}_{it} 
\end{equation}

\noindent where $\delta^{b}_{it}$ includes the unobserved (to researchers) traffic-specific behavioural component common to all drivers, traffic-specific vehicular attribute common to all vehicles, weather-specific component affecting the entire traffic stream, and demand-related characteristic. $\xi^{b}_{it}$ represents an idiosyncratic error term representing all random shocks to the dependent variable. Since the exact structural form of how $o^b_{it}$ enters into equation is unknown, we adopt a non-parametric specification of $S^{b}(.)$. $\delta^{b}_{it}$ is expected to be correlated with $o^b_{it}$. We explain the implications of this correlation on the estimated relationship in the next subsection (Section \ref{S:4.2}). 

As a by-product of this estimation, we quantify the activation of the bottleneck as follows. Consistent with the engineering literature, we consider that the flow through the bottleneck drops following the activation of the bottleneck. Thus, we infer the critical value of $o^b_{c}$ at which we observe a significant backward bending in $q^b$ from the estimated relationship $S^{b}(.)$. We also note that when the occupancy inside the bottleneck remains at and above $o^b_{c}$, the bottleneck remains activated.

Through the estimated relationship, we quantify the capacity of the bottleneck $q^b_{c}$, that, is flow through the bottleneck corresponding to $o^b_{c}$ and examine the existence of capacity drop or two-capacity phenomenon following the activation of the bottleneck. 

\subsection{Bias due to Endogeneity}
\label{S:4.2}

Building a credible causal relationship between traffic variables requires the understanding of potential endogeneity biases. There are two major concerns in relation to endogeneity: omitted variable bias and reverse causality (simultaneity). Omitted covariates that are correlated with both the dependent variable and the included covariates in a regression may result in inconsistent estimates of model parameters. For instance, in equation \ref{eq:spec}, omission of covariates representing driving behaviour of users due to unavailability of a comprehensive aggregate level measure may lead to confounding bias in the estimated relationship. This bias occurs because driving behaviour may be correlated with both occupancy and flow. Reverse causality is a consequence of the existence of a two-way causal relationship or a cause-effect relationship, contrary to the one assumed in the model. For instance, in equation \ref{eq:spec}, we assume the flow through the bottleneck, $q^b_{it}$ to be a function of the occupancy inside the bottleneck, $o^b_{it}$. However, there may be reverse causality where $q^b_{it}$ affects $o^b_{it}$ in certain traffic situations. The presence of reverse causality may also lead to inconsistent estimates. We further mathematically demonstrate the two sources of confounding and resulting biases in Appendix A.

\subsection{Bayesian Nonparametric Instrumental Variable Approach}
\label{S:4.3}

To address both endogeneity biases, we adopt regression estimators with instrumental variables (IV). IV-based estimators such as two-stage least squares (2SLS) are widely adopted in applied econometrics to estimate parametric models that contain endogenous explanatory variables \citep[see, for example][]{Wooldridge2010}. However,  finite-dimensional parametric models for the fundamental relationship of traffic flow such as a linear speed-density or a quadratic flow-speed model, are based on assumptions that are rarely justified by engineering or economic theories. The resulting model mis-specification may lead to erroneous estimates of attributes characterising the fundamental relationship (for instance, capacity or capacity drop). On the other hand, non-parametric methods have the potential to capture the salient features in a data-driven manner without making a priori assumptions on the functional form of the relationship \citep{Horowitz2011}. Therefore, a fairly growing strand in the econometrics literature proposes different approaches for non-parametric instrumental variables (NPIV) regression, but such methods have not been considered in the estimation of fundamental diagram. Extensive reviews can be found in \cite{Newey2003} and \cite{Horowitz2011}. 
The NPIV approaches are either based on regularisation or control function. In this study, we adopt a control-function based Bayesian NPIV estimator \citep{Wisenfarth2014}. In what follows, we start with the general model set-up. Subsequently, we discuss the advantages of the adopted control-function-based Bayesian approach. Additionally, in Appendix B, we summarise the challenges associated with regularisation based approaches that are more commonly adopted in the empirical economics literature. 

We first rewrite equation \ref{eq:spec} in a traditional two-stage IV-based regression set up: 

\begin{equation}
\label{eq:NPIV1}
q = S(o) + \epsilon_2, \quad o = h(z) + \epsilon_1
\end{equation}

\noindent with response $q$, endogenous covariate $o$, IV $z$ for $o$ and idiosyncratic error terms $\epsilon_1$ and $\epsilon_2$ for the first and second stage regressions, respectively. For the notational simplicity, we drop time-day subscripts and superscripts. Note that $\delta$ in equation \ref{eq:spec} are encapsulated in $\epsilon_2$. Endogeneity bias arises as $E(\epsilon_2|o)\neq 0$. We assume the following identification restrictions: 

\begin{equation}
\label{eq:NPIV2}
E(\epsilon_1|z) = 0 \quad and \quad E(\epsilon_2|\epsilon_1,z) = E(\epsilon_2|\epsilon_1),
\end{equation}

\noindent which yields

\begin{equation}
\label{eq:NPIV3}
\begin{split}
E(q|o,z) & = S(o) + E(\epsilon_2|\epsilon_1,z) = S(o) + E(\epsilon_2|\epsilon_1) \\
             & = S(o) + \nu(\epsilon_1),
\end{split}
\end{equation}

\noindent where $\nu(\epsilon_1)$ is a function of the unobserved error term $\epsilon_1$. This function is known as the control function. 

\subsubsection{Control function-based approaches}

Several control function-based approaches to estimation of equation \ref{eq:NPIV1} in the literature, adopt a two-stage approach where residuals $\hat{\epsilon_1}$, that is, $o-\hat{h(z)}$ from the first stage are used as additional covariate in the second stage \citep[for details, see][]{Newey2003}. However, as pointed out by \cite{Wisenfarth2014}, such two-stage approaches have certain limitations. First, the uncertainty introduced by estimating the parameters in the first stage remains unincorporated in the second stage. Second, a precise estimate of $\nu(\epsilon_1)$ to achieve full control for endogeneity is difficult to obtain because the focus is on minimising the error in predicting $o$ in the first stage. Third, a robustness control is required to account for outliers and extreme observations in $\epsilon_1$ that may affect the endogeneity correction. 

Bayesian control-function-based approaches can address these shortcomings of frequentist counterparts and regularisation-based approaches by estimating equation \ref{eq:NPIV1} as a simultaneous system of equations, allowing for automatic smoothing parameter selection for a precise estimation of the control function and for construction of simultaneous credible bands \footnote{Credible bands are the Bayesian analogue to confidence bands in the frequentist set up that represent the uncertainty of an estimated curve.}. 

However, early Bayesian control-function-based approaches consider a bivariate Gaussian distribution of errors $(\epsilon_1,\epsilon_2) \sim N(0,\Sigma)$ \citep[for instance, see][]{Chib2009}. This assumption leads to linearity of the conditional expectation as, $E(\epsilon_1|\epsilon_2) = \frac{\sigma_{12}}{\sigma_1^2}$, where $\sigma_{12}=cov(\epsilon_{1},\epsilon_{2})$ and $\sigma_1^2=var(\epsilon_{1})$, restricting the control function to be linear in $\epsilon_1$ \citep{Wisenfarth2014,Conley2008}.  Since outliers can be a common source of non-linearity in error terms, they can aggravate the robustness issues of such linear specifications. To overcome these limitations, \cite{Conley2008} proposed the application of a Dirichlet process mixture (DPM) prior to obtain a flexible error distribution, but still relied on linear covariate effects. The method proposed by \cite{Wisenfarth2014} and adopted in this study, extends the approach by \cite{Conley2008} and allows for fully-flexible covariate effects. 

\subsubsection{Adopted Bayesian NPIV approach \citep{Wisenfarth2014}}

The \cite{Wisenfarth2014}'s Bayesian NPIV approach thus allows us to correct for endogeneity bias in regression models where the covariate effects and error distributions are learned in a data-driven manner, obviating the need of a priori assumptions on the functional form.

To satisfy the identification restrictions presented in equation \ref{eq:NPIV2}, we need an instrumental variable (IV) $z$. The IV should be (i) exogenous, that is, uncorrelated with $\epsilon_2$; (ii) relevant, that is, correlated with the endogenous covariate $o$, conditional on other covariates in the model. Due to the absence of suitable external instruments, we use an aggregate lagged level of the endogenous covariate (occupancy) as an instrument. Specifically, for occupancy observed in the five-minutes interval $i$ on day $t$, we consider the average of observations on the covariate from the interval $i-15$ to $i+15$ from the previous workday $t-1$ as its instrument. We argue that the occupancy $o_{it}^b$ in the five-minutes interval $i$ on day $t$ is correlated with the occupancy $o_{[i-15,i+15],t-1}^b$ in the thirty-minutes interval surrounding $i$ on the previous day $t-1$. This correlation follows from the influence of time-of-the-day on demand and from the fact that there are no reasonable alternative routes to the highway sections being studied, so the population of drivers using the section in the duration $i$ over different workdays may not differ substantially. Moreover, as the highway infrastructure remains unaltered during the study period, we expect the average driving behaviour and thus traffic density to not differ substantially over different days as the drivers are already conversant with the route. However, these lagged occupancy values $o_{[i-15,i+15],t-1}^b$ are exogenous because they do not directly determine the response variable $q_{it}^b$ in equation \ref{eq:spec} and would never feature in the model for that response. To justify the relevance of the considered instrument, we present the estimated $h(.)$ in equation \ref{eq:NPIV1} and complimentary results from \cite{Stock2005} weak instrument F-tests in the Results and Discussion Section (Section \ref{S:5.2.2}).

Conditional on the availability of an instrument, the Bayesian NPIV estimator can  correct for the confounding bias. To account for nonlinear effects of continuous covariates, both $S(.)$ and $h(.)$ (refer equation \ref{eq:NPIV1}) are specified in terms of additive predictors comprising penalised splines. Each of the functions $S()$ and $h(.)$ is approximated by a linear combination of suitable B-spline basis functions. The penalised spline approach uses a large enough number of equidistant knots in combination with a penalty to avoid over-fitting. Moreover, the joint distribution of $\epsilon_1$ and $\epsilon_2$ is specified using nonparametric Gaussian DPM, which ensures robustness of the model relative to extreme observations. Efficient Markov chain Monte Carlo (MCMC) simulation technique is employed for a fully Bayesian inference. The resulting posterior samples allow us to construct simultaneous credible bands for the non-parametric effects (i.e., $S(.)$ and $h(.)$). Thereby, the possibility of non-normal error distribution is considered and the complete variability is represented by Bayesian NPIV. We now succinctly discuss specifications of the kernel error distribution and computation of credible bands in Bayesian NPIV. 

To allow for a flexible distribution of error terms, the model considers a Gaussian DPM with infinite mixture components, $c$, in the following hierarchy:

\begin{equation}
\label{eq:NPIV_error}
\begin{gathered}
(\epsilon_{1i},\epsilon_{2i}) \: \sim  \: \sum_{c=1}^\infty \pi_c \textrm{N}(\mu_c,\Sigma_c) \\
(\mu_c,\Sigma_c) \: \sim  \: G_0 = \textrm{N}(\mu|\mu_0, \tau_{\Sigma}^{-1}\Sigma) \: \textrm{IW}(\Sigma|s_\Sigma,S_\Sigma)\\
\pi_c=\upsilon_c\left(1-\sum_{j=1}^{c-1}(1-\pi_j)\right)=\upsilon_c\prod_{j=1}^{c-1}(1-\upsilon_j),\\
c=1,2,...\\
\upsilon_c  \: \sim \: \textrm{Be}(1,\zeta).\\
\end{gathered}
\end{equation}

\noindent where $\mu_c$, $\Sigma_c$ and $\pi_c$ denote the component-specific means, variances and mixing proportions. The mixture components are assumed to be independent and identically distributed with the base distribution $G_0$ of the Dirichlet process (DP), where $G_0$ is given by a normal-inverse-Wishart distribution. The mixture weights are generated in a stick-breaking manner based on a Beta distribution with concentration parameter $\zeta>0$ of the DP. The concentration parameter $\zeta$ determines the strength of belief in the base distribution $G_0$.

\subsubsection{Estimation Practicalities}

We exclude Gibbs sampler of Bayesian NPIV for brevity, but mainly focus on implementation details and posterior analysis. Interested readers can refer to \cite{Wisenfarth2014} for derivation of conditional posterior updates. 

We adapt \textit{BayesIV} and \textit{DPpackage} in R to estimate Bayesian NPIV. We consider 50,000 posterior draws in the estimation, exclude the first 15,000 burn-in draws and keep every 10$^{th}$ draw from the remaining draws for the posterior analysis. The point-wise posterior mean is computed by taking the average of 3,500 posterior draws. Bayesian simultaneous credible bands are obtained using quantiles of the posterior draws. A simultaneous credible band is defined as the region $I_\delta$ such that $P_{S|data}(S \in I_\delta) = 1-\delta$, that is, the posterior probability that the entire true function $S(.)$ is inside the region given the data equals to $1-\delta$. The Bayesian simultaneous credible bands are constructed using the point-wise credible intervals derived from the $\delta/2$ and $1-\delta/2$ quantiles of the posterior samples of $S(.)$ from the MCMC output such that $(1-\delta)100\%$ of the sampled curves are contained in the credible band. Similar process is used to obtain the credible intervals of $h(.)$.

\subsection{Monte Carlo Simulations}
\label{S:4.4}

We succinctly demonstrate the ability of the adopted Bayesian NPIV approach in addressing challenges of functional form mis-specification and endogeneity in an instance of a Monte Carlo study. We benchmark the Bayesian NPIV method against state-of-the-art estimators to illustrate how this method is robust to both issues. In the data generating process (DGP), we consider a concave regression function, that is, a fourth degree polynomial specification but our conclusions are applicable for a more complex specification.

We use a sample of 10000 observations, with the following DGP:

\begin{equation}
\label{eq:sim_1}
\begin{gathered}
y = -40 x^4 + 40 x^3 + 30 w^4 + \epsilon_2 \\
x = 3.5 z + 2.1 w + \epsilon_1 \\
\end{gathered}
\end{equation}

\noindent where $z$ and $w$ are independent and uniformly distributed on [0,1]. $\epsilon_1$ and $\epsilon_2$ are independent and identically distributed draws from the N [0,0.5] distribution. The variables $y$ represents the primary response variable, $x$ denotes the endogenous covariate and $z$ represents the instrumental variable. The variable $w$ captures the unobserved effects in the model, that is, we assume that the analyst is ignorant of its presence in the true data generating process for the dependent variable. We thus introduce one possible source of confounding into the model: a positive correlation between the unobserved effect $w$ and the endogenous covariate $x$.

We note that the model set-up is similar in structure to equation \ref{eq:NPIV1}, that is:

\begin{equation}
\label{eq:sim_2}
\begin{gathered}
y = s(x) + \epsilon_2 \\
x = h(z) + \epsilon_1 \\
\end{gathered}
\end{equation}

We apply four different estimators to estimate the curve $s(x)$: 

\begin{enumerate}
    \item A two-stage least square (2SLS) estimator with a quadratic specification for $s(x)$.\footnote{Instead of a traditionally-used linear specification, we choose quadratic specification in 2SLS because the scatter plot of the data would intuitively suggest the analyst to use such functional form of $s(x)$.}
    \item A two-stage least square (2SLS) estimator with the true specification for $s(x)$.
    \item A Bayesian non-parametric estimator without instrumental variables (Bayes NP).
    \item A Bayesian non-parametric estimator with instrumental variables (Bayes NPIV).
\end{enumerate}

\begin{figure}[t]
  \centering
  \includegraphics[width= 0.9\linewidth]{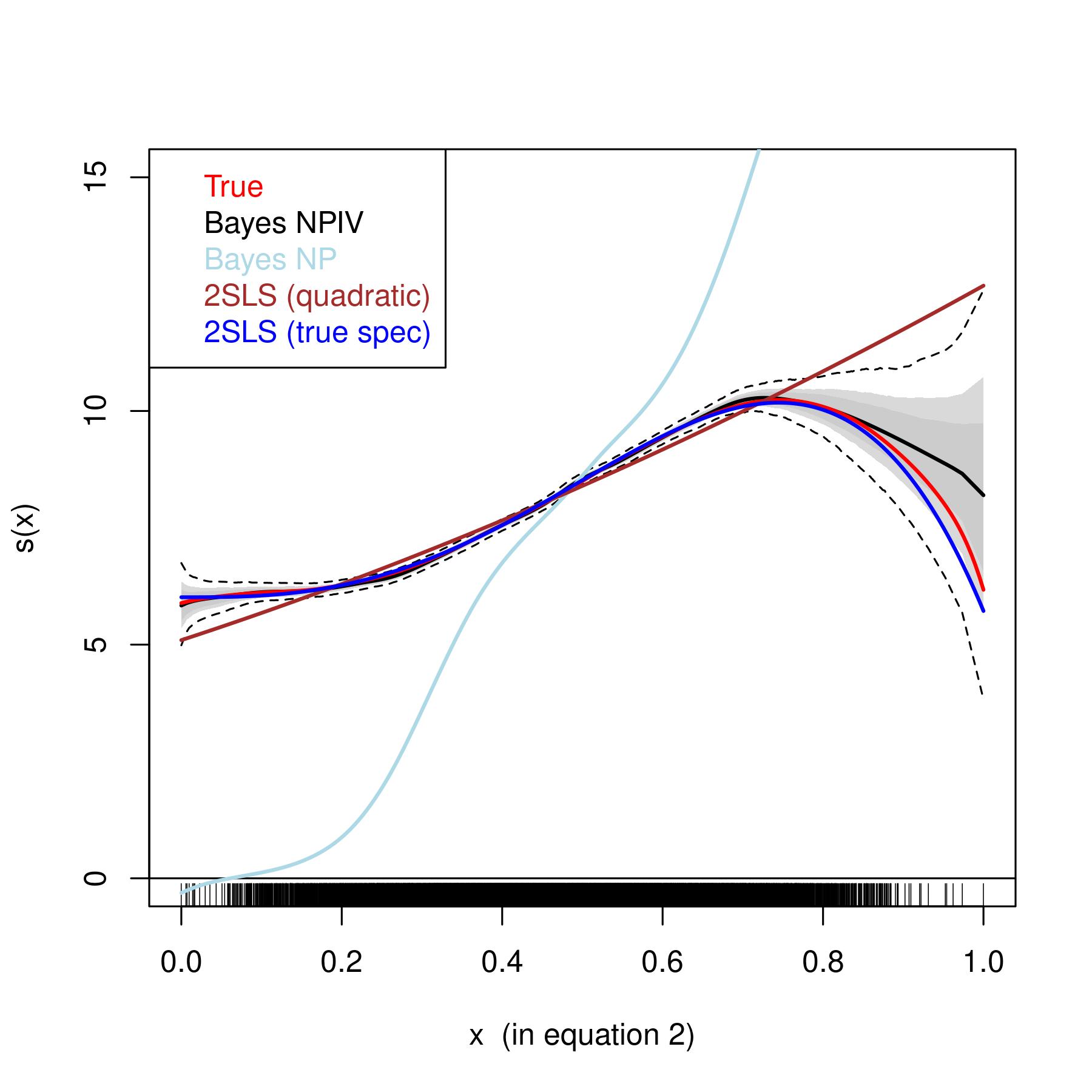}
  \caption{Comparison of different estimators in the Monte Carlo study.}
  \label{fig:sim}
\end{figure}

In the latter two approaches, we take 40000 Posterior draws to ensure stationarity of Markov chains. For the posterior analysis, the initial 10000 draws were discarded for burn-in and every 40th draw of the subsequent 30,000 draws was used for posterior inference. Figure \ref{fig:sim} overlays the estimated $s(x)$ from the four approaches and true $s(x)$. 

We note that a 2SLS estimator with the true specification for $s(x)$ is able adjust for the endogeneity bias and could produce an unbiased estimate of $s(x)$. However, in practice, it is infeasible for the analyst to know the correct functional form specification a priori. A functional form mis-specification can produce a highly biased estimate of $s(x)$, as shown by the estimated $s(x)$ using the 2SLS estimator with a quadratic specification for $s(x)$. This exercise thus illustrates the importance of adopting a fully flexible non-parametric specification for $s(x)$ in a relationship.

However, in the presence of endogeneity, a traditional non-parametric estimator may fail to produce an unbiased estimate of $s(x)$.  From Figure \ref{fig:sim}, we note that the curve produced by the Bayes NP is highly biased. Adopting an estimator such as the Bayes NPIV allows to adjusts for the endogeneity bias and produce an unbiased estimate of the curve $s(x)$. 

In summary, this Monte Carlo exercise shows that the Bayes NPIV estimator, the one adopted in this study, outperforms other parametric and non-parametric approaches as it is allows for a fully flexible functional form specification and controls for any potential confounding bias.

\section{Results and Discussion}
\label{S:5}

This section is divided into two subsections. In the first subsection, we compare results of the adopted Bayesian NPIV estimator with those of a Bayesian NP estimator and a pooled ordinary least squares (POLS) estimator with a quadratic specification. The Bayesian NP estimator is a counterpart of the Bayesian NPIV, which does not address confounding bias (that is, $z=x; \epsilon_1 = 0; h(.): \textrm{identity function}$ in Equation \ref{eq:NPIV1}). Furthermore, we discuss the estimates of the capacity and capacity-drop in detail and compare these values with those reported in the literature. In the next subsection, we present the estimated kernel error distributions to illustrate the importance of the non-parametric DPM specification. The relevance of our instruments is also demonstrated in this subsection.
 
\subsection{Comparison of Bayesian NPIV and non-IV-based estimators}
\label{S:5.1}

\begin{figure}[tp]
    \centering
        \begin{subfigure}{0.5\textwidth}
            \centering
            \includegraphics[width=1\textwidth]{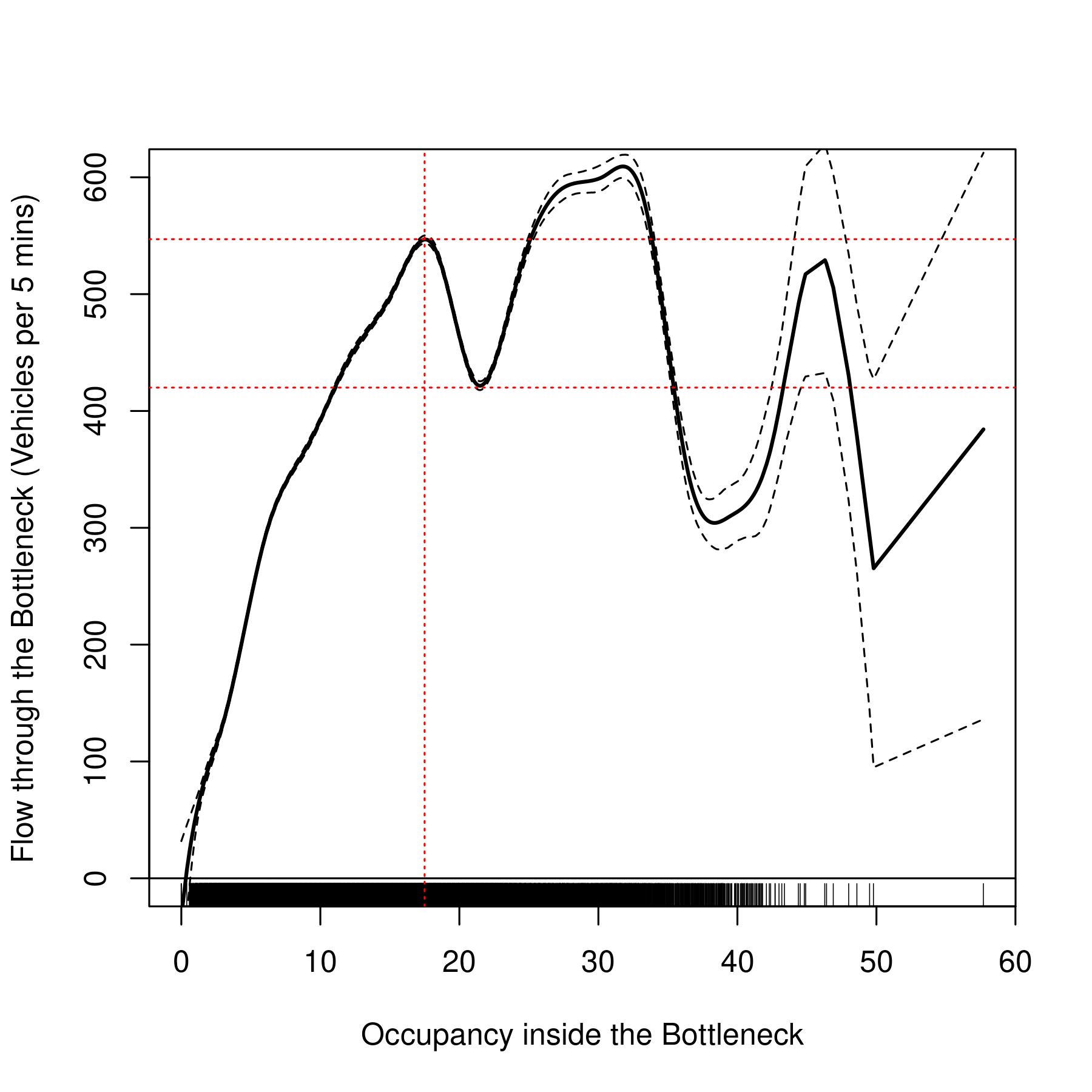}
            \caption{Non-parametric non-IV estimator.}
        \end{subfigure}%
        \hfill
        \begin{subfigure}{0.5\textwidth}
            \centering
            \includegraphics[width=1\textwidth]{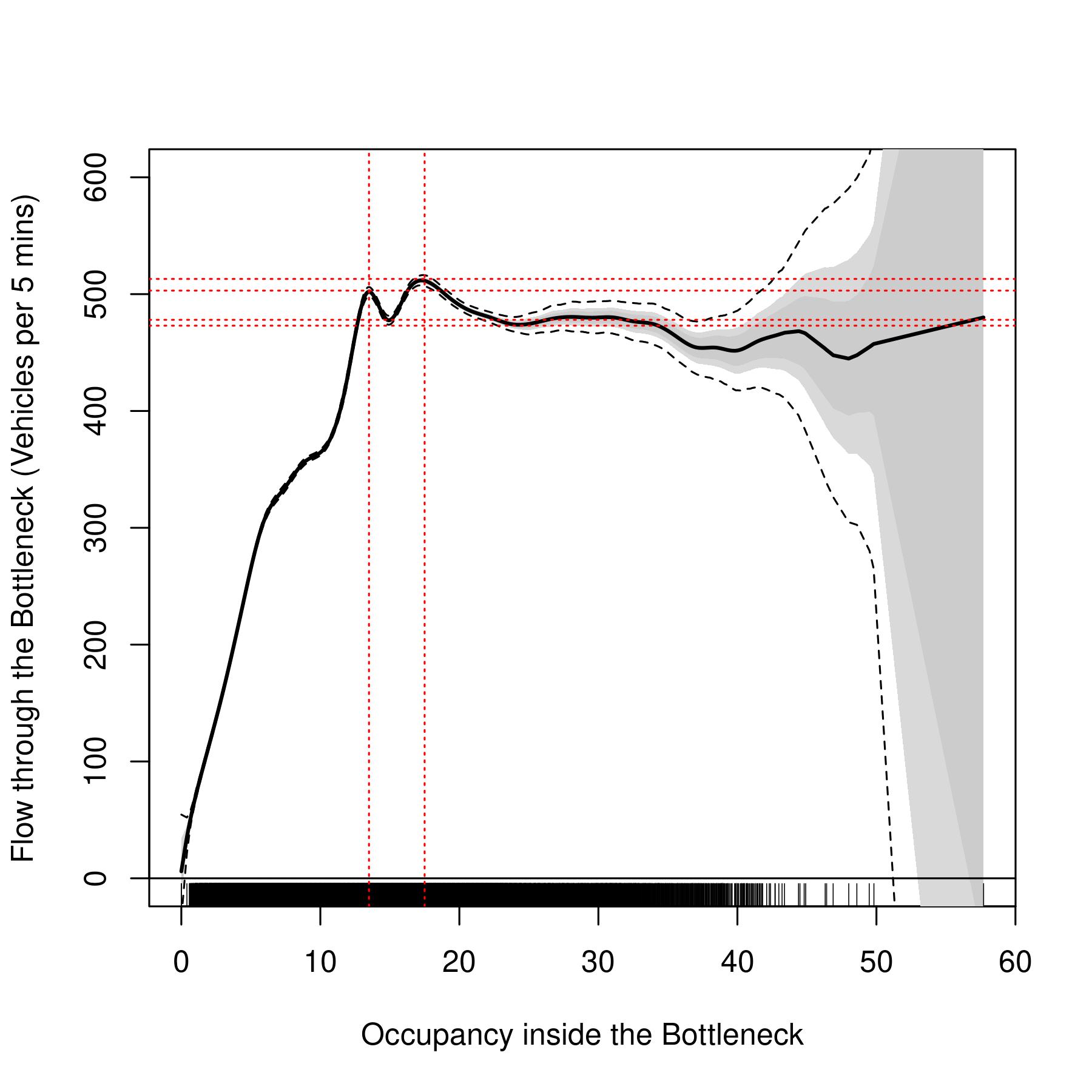}
            \caption{Non-parametric IV-based estimator.}
        \end{subfigure}
        \hfill
        \begin{subfigure}{0.8\textwidth}
            \centering
            \includegraphics[width=1\textwidth]{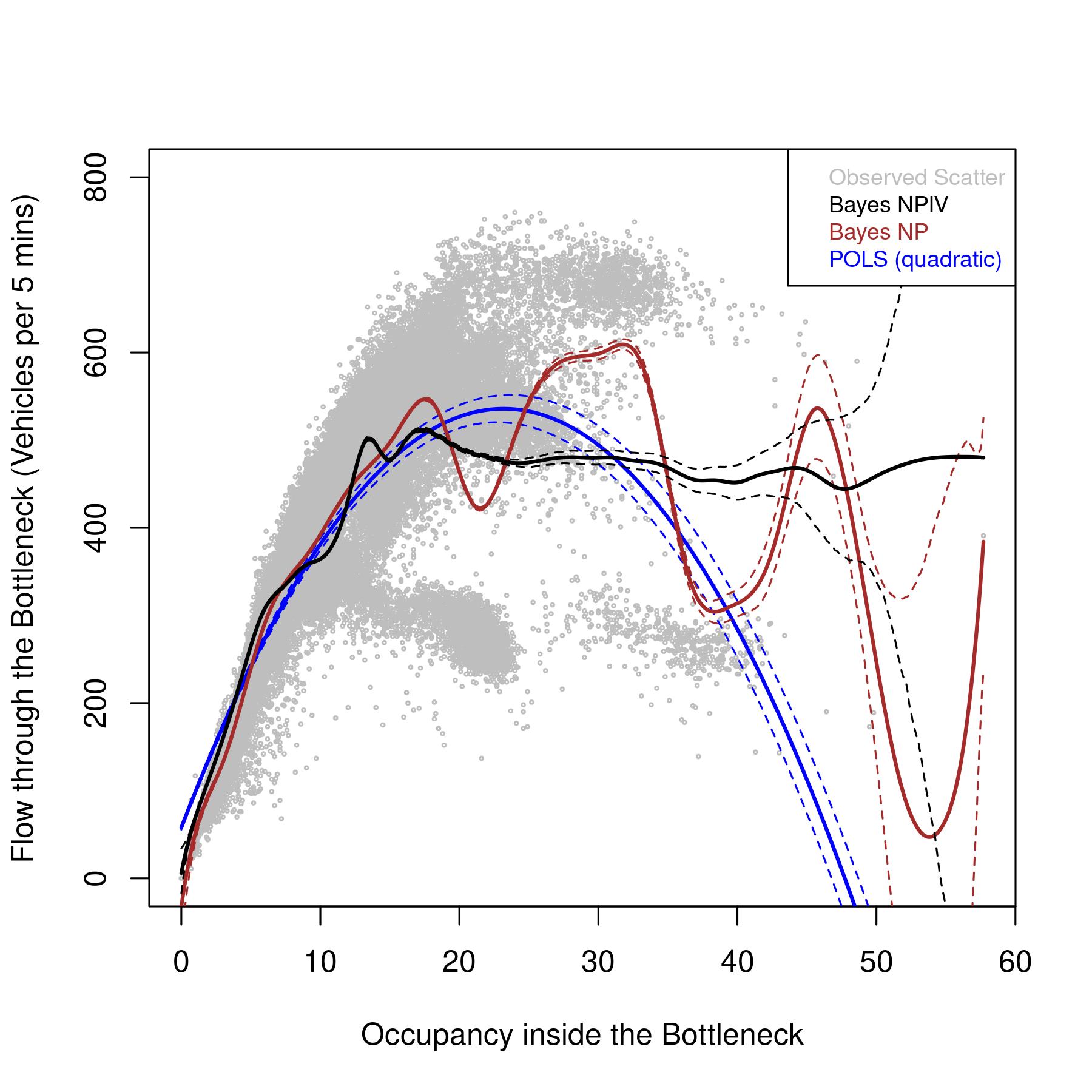}
            \caption{Comparison of different estimators.}
        \end{subfigure}
    \caption{Estimated flow-occupancy curves for Westbound SR-24.}
    \label{fig:SR24W}
\end{figure}

\begin{figure}[tp]
    \centering
        \begin{subfigure}{0.5\textwidth}
            \centering
            \includegraphics[width=1\textwidth]{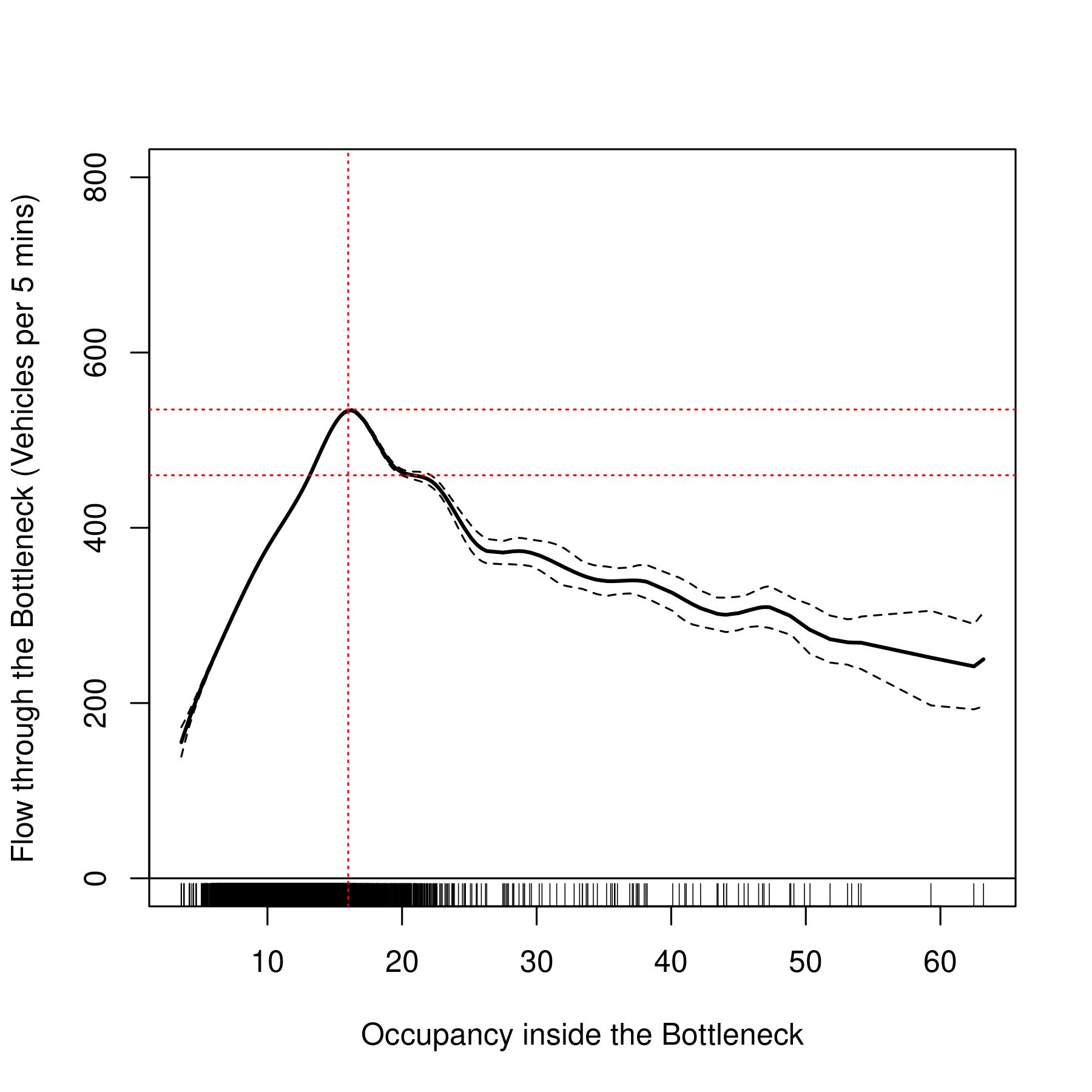}
            \caption{Non-parametric non-IV estimator.}
        \end{subfigure}%
        \hfill
        \begin{subfigure}{0.5\textwidth}
            \centering
            \includegraphics[width=1\textwidth]{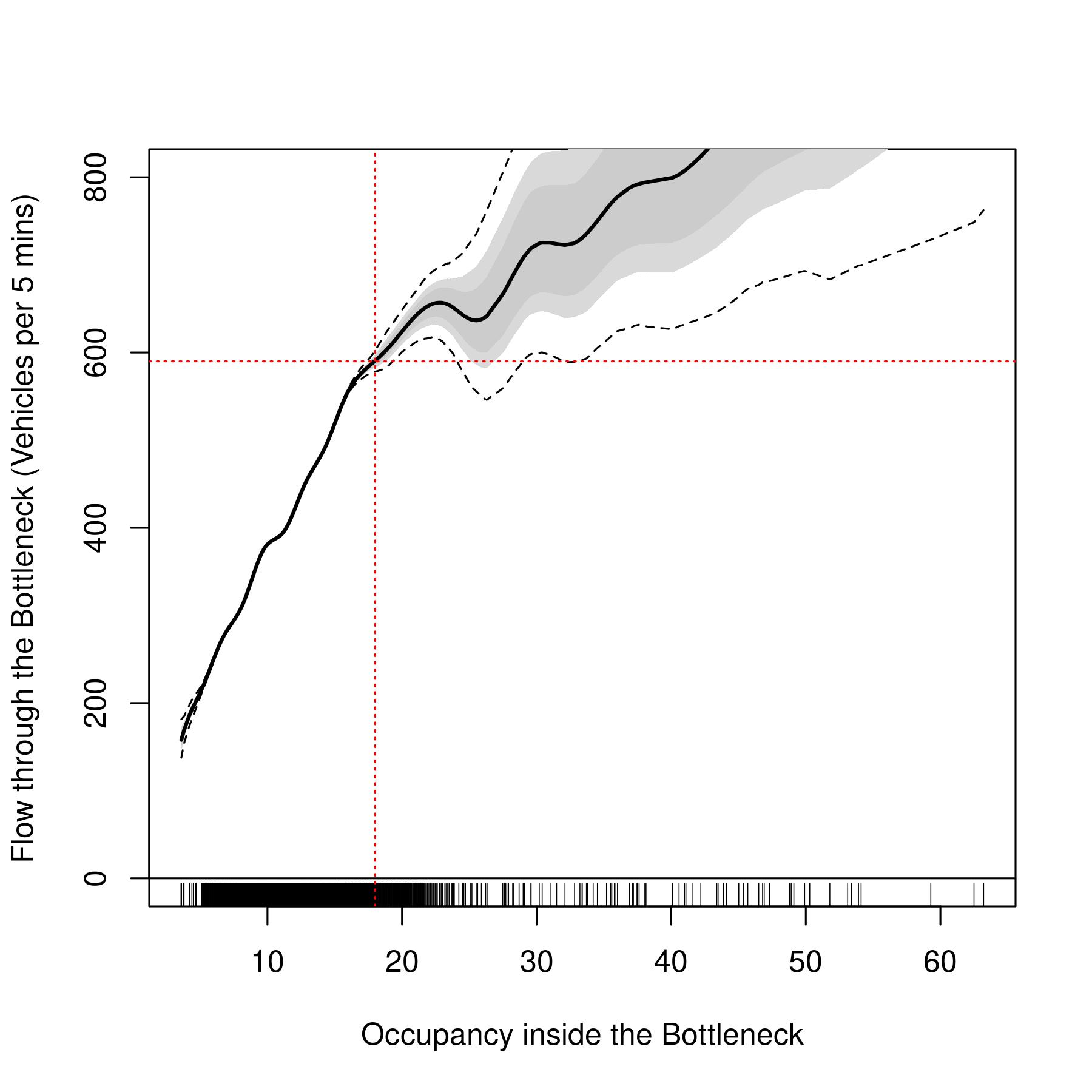}
            \caption{Non-parametric IV-based estimator.}
        \end{subfigure}
        \hfill
        \begin{subfigure}{0.8\textwidth}
            \centering
            \includegraphics[width=1\textwidth]{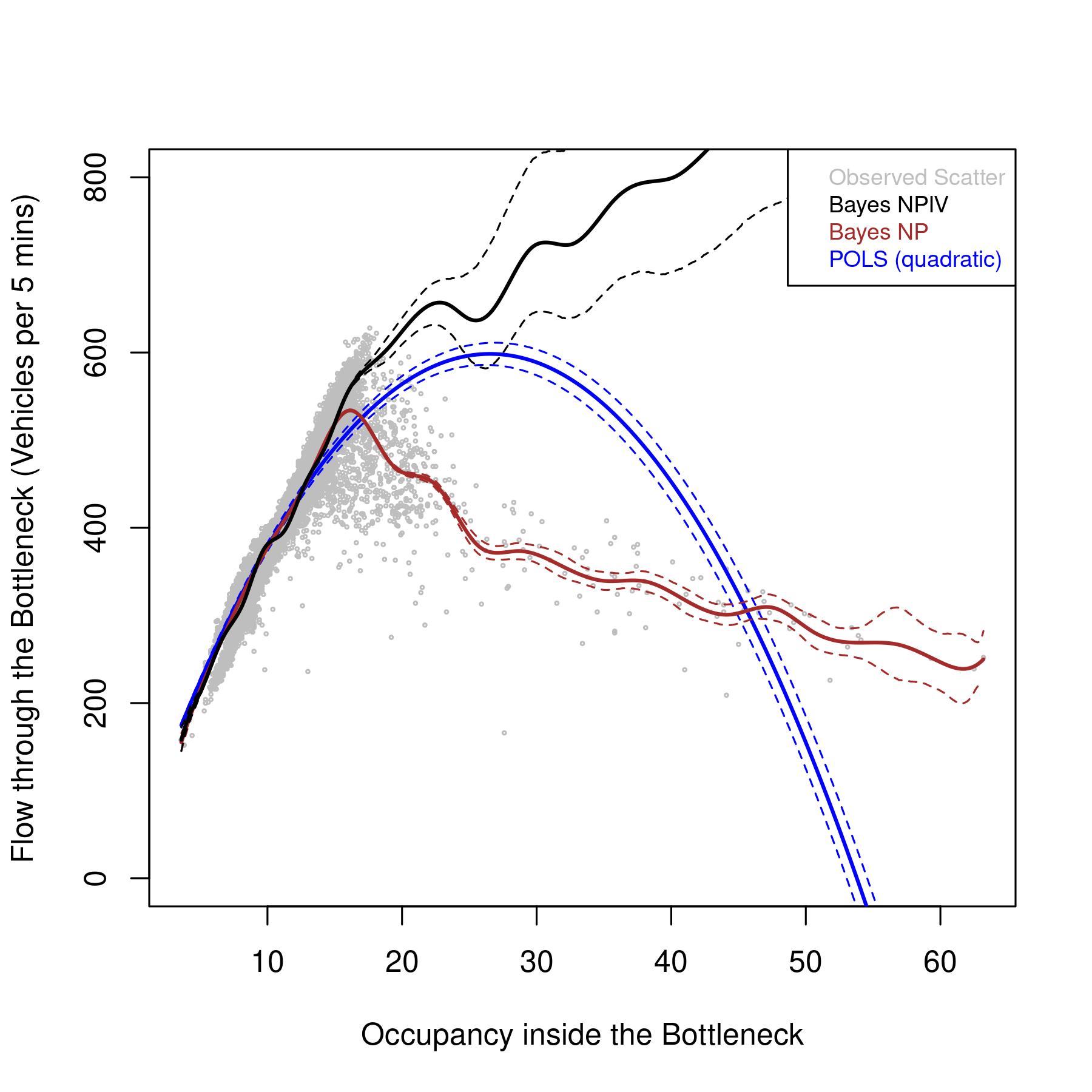}
            \caption{Comparison of different estimators.}
        \end{subfigure}
    \caption{Estimated flow-occupancy curves for Eastbound SR-91.}
    \label{fig:SR91E}
\end{figure}

\begin{figure}[tp]
    \centering
        \begin{subfigure}{0.5\textwidth}
            \centering
            \includegraphics[width=1\textwidth]{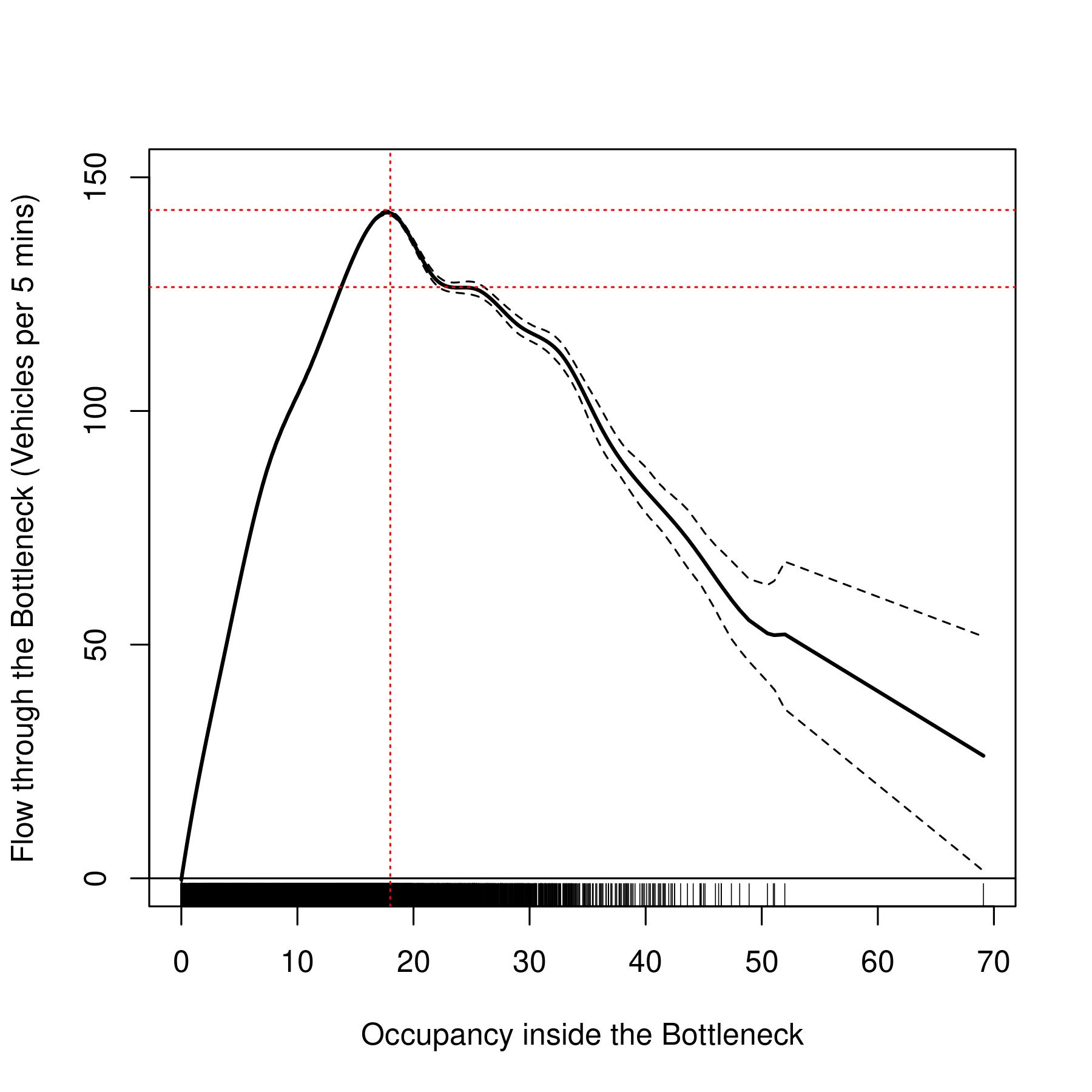}
            \caption{Non-parametric non-IV estimator.}
        \end{subfigure}%
        \hfill
        \begin{subfigure}{0.5\textwidth}
            \centering
            \includegraphics[width=1\textwidth]{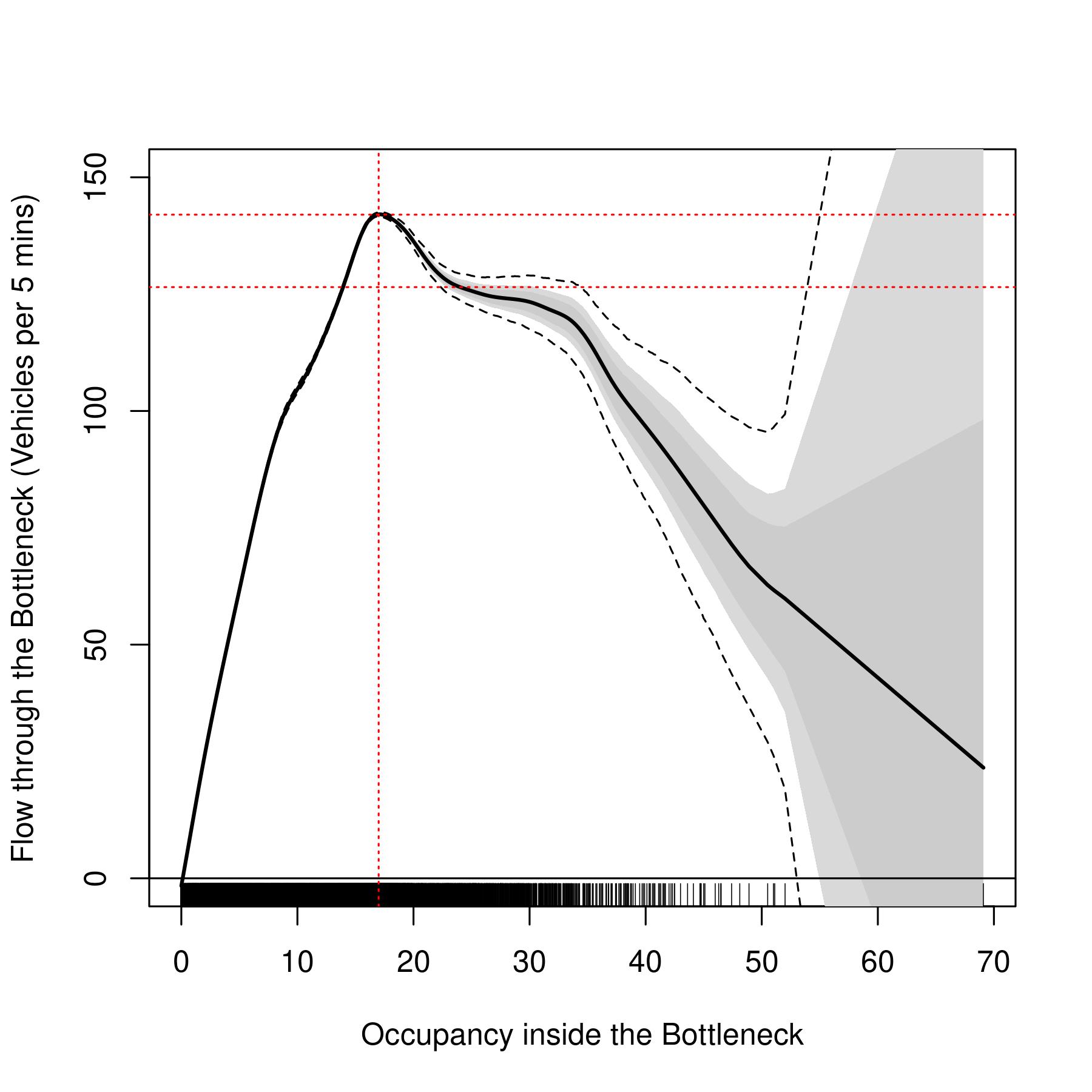}
            \caption{Non-parametric IV-based estimator.}
        \end{subfigure}
        \hfill
        \begin{subfigure}{0.8\textwidth}
            \centering
            \includegraphics[width=1\textwidth]{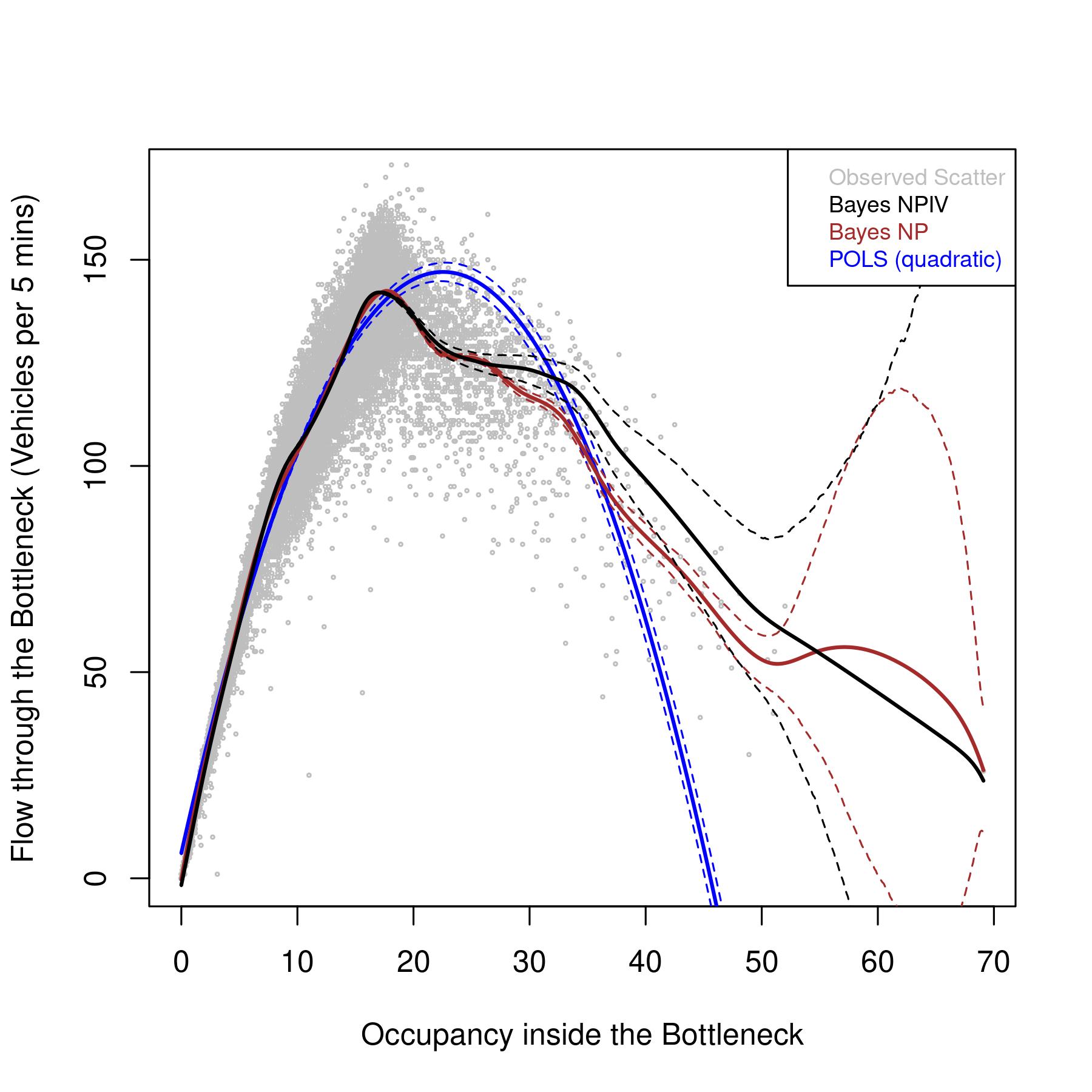}
            \caption{Comparison of different estimators.}
        \end{subfigure}
    \caption{Estimated flow-occupancy curves for Eastbound SR-12.}
    \label{fig:SR12E}
\end{figure}

We present the estimates of $S(.)$ (see equation \ref{eq:NPIV1}, second-stage) using Bayesian NPIV, Bayesian NP, and POLS in Figures \ref{fig:SR24W}, \ref{fig:SR91E} and \ref{fig:SR12E} for the three highway sections. POLS results are mainly presented to illustrate how commonly-used parametric non-IV-based specifications can result biased results, but most discussion would revolve around comparing results of Bayesian NPIV and its non-IV counter part (that is, Bayesian NP).   

From each of these figures, we do not observe any notable differences between the Bayesian NPIV and Bayesian NP estimate of the free-flow regime of the flow-occupancy curve. In this regime, the Bayesian NPIV estimate of $S(.)$ is as efficient as its Bayesian NP counterpart, as evidenced by tight credible bands in the domain of occupancy where we have sufficient number of observations (note that the density of the tick marks on the X-axis represents the number of observations). However, we observe substantial differences near the saturation (capacity) point and in the congested (or hypercongested as per the economics literature) regime of the estimate curve (see Figures \ref{fig:SR24W}(c), \ref{fig:SR91E}(c) and \ref{fig:SR12E}(c)). We further discuss these differences in detail in next sub-sections.

\begin{table}[tp]
\caption{Summary of Results.}
\label{tab:Results}
\centering
\begin{subtable}[c]{1\textwidth}
\caption{Comparison of estimators.}
\label{tab:Comp}
\centering
\begin{adjustbox}{width= 1\textwidth}
\begin{tabular}{L{3.5cm} C{2.5cm} C{2.5cm} C{2.3cm} C{2.3cm} } 
\hline
& \multicolumn{2}{c}{Estimated Capacity (veh/hr)} & \multicolumn{2}{c}{Estimated Capacity-drop (percent)} \\
Highway Section & Bayes NP & Bayes NPIV & Bayes NP & Bayes NPIV\\
\hline
Westbound SR-24 & 6561.12 (12.00) & 6141.00 (15.24) & 27.42 & 7.80 \\ 
Eastbound SR-91 & 6407.52 (5.16) & 6997.44 (30.96) & 14.02 & n.s.\\ 
Eastbound SR-12 & 1709.40 (1.44) & 1708.20 (1.68) & 11.54 & 10.92 \\ 
\hline
\multicolumn{5}{l}{\footnotesize{*n.s. stands for not statistically significant.}}\\
\multicolumn{5}{l}{\footnotesize{**Figures in bracket indicate the associated standard errors.}}
\end{tabular}
\end{adjustbox}
\bigskip
\end{subtable}
\hfill
\begin{subtable}[c]{\textwidth}
\caption{Estimated capacity and comparison with the literature.}
\label{tab:Cap}
\centering
\begin{adjustbox}{width= 1\textwidth}
\begin{tabular}{L{3.5cm} C{4cm} C{9cm}} 
\hline
Highway Section & Estimated Capacity & Capacity reported in the Engineering literature\\
\hline
Westbound SR-24 & 6141 veh/hr & ~4100 veh/hr \citep{Chung2007} \\ 
Eastbound SR-91 & 6997 veh/hr & ~7200 veh/hr \citep{Oh2012} \\ 
Eastbound SR-12 & 1708 veh/hr & NA \\ 
\hline
\multicolumn{3}{l}{\footnotesize{*NA stands for not available.}}\\
\multicolumn{3}{l}{\footnotesize{**Figures in bracket indicate the associated standard errors.}}
\end{tabular}
\end{adjustbox}
\bigskip
\end{subtable}
\hfill
\begin{subtable}[c]{\textwidth}
\caption{Estimated capacity-drop and comparison with the literature.}
\label{tab:Cap_drop}
\centering
\begin{adjustbox}{width= 1\textwidth}
\begin{tabular}{L{3.5cm} C{3cm} C{4cm} C{4cm}} 
\hline
& Estimated & \multicolumn{2}{c}{Average Capacity-drop as reported in the} \\
Highway Section & Capacity-drop & Engineering literature & Economics literature\\
\hline
Westbound SR-24 & 7.80 percent & 5.10 to 8.40 percent &  n.s. \\ 
Eastbound SR-91 & n.s. & 13.50 percent &  NA \\ 
Eastbound SR-12 & 10.92 percent & NA &  n.s. \\ 
\hline
\multicolumn{4}{l}{\footnotesize{*n.s. stands for not statistically significant; NA stands for not available.}}\'\
\end{tabular}
\end{adjustbox}
\bigskip
\end{subtable}
\hfill
\begin{subtable}[c]{\textwidth}
\caption{Activation of the bottleneck.}
\label{tab:Occ}
\centering
\begin{adjustbox}{width= 0.65\textwidth}
\begin{tabular}{L{3.5cm} C{3.0 cm} C{3.0cm}} 
\hline
& \multicolumn{2}{c}{Occupancy corresponding to capacity} \\
Highway Section & non-IV-based & IV-based \\
\hline
Westbound SR-24 & 17.56 & 17.27 \\ 
Eastbound SR-91 & 16.13 & 17.50 \\ 
Eastbound SR-12 & 17.71 & 17.08 \\ 
\hline
\end{tabular}
\end{adjustbox}
\bigskip
\end{subtable}
\end{table}

\newpage

\subsubsection{Estimated capacity}
\label{S:5.1.1}

Table \ref{tab:Comp} summarises the estimated capacity for each highway section. For Westbound SR-24 that features a lane-drop bottleneck with number of lanes reducing from four to two, the capacity estimated via the Bayesian NP estimator, that is, 546.76 (1.00) vehicles per five-minutes or 6561.12 (12.00) vehicles per hour, is significantly more that the Bayesian NPIV-based estimate, that is, 511.75 (1.27) vehicles per five-minutes or 6141.00 (15.24) vehicles per hour (see Figure \ref{fig:SR24W}). The capacity reported in the engineering literature is 4100 vehicles per hour (refer to Table \ref{tab:Cap}), which is much lower than both of these estimates. 

For Eastbound SR-91 that features a merge bottleneck, Bayesian NPIV-based estimate of capacity is 583.12 (2.58) vehicles per five-minutes or 6997.44 (30.96) vehicles per hour, which is significantly higher that the Bayesian NP-based estimate of 533.96 (0.43) vehicles per five-minutes or 6407.52 (5.16) vehicles per hour (see Figure \ref{fig:SR91E}) but is consistent with the value reported in the engineering literature (7200 vehicles per hour, see Table \ref{tab:Cap}).

For Eastbound SR-12 that features a lane-drop bottleneck with number of lanes reducing from two to one, the capacity estimated via the Bayesian NP-based estimator, that is, 142.45 (0.12) vehicles per five-minutes or 1709.40 (1.44) vehicles per hour, is similar to the Bayesian NPIV-based estimate of 142.35 (0.14) vehicles per five-minutes or 1708.20 (1.68) vehicles per hour (see Figure \ref{fig:SR12E}). We do not note any previous estimate of capacity of this section from the literature.

The above comparison does not point towards a clear direction of bias in the Bayesian NP-based estimate of capacity with respect to the Bayesian NPIV-based estimate, rather it varies on a case-by-case basis depending upon the data generating process. Failing to address endogeneity bias leads to an over-estimation, an under-estimation and no difference in the estimated capacity for the first, second and third sections, respectively. We also find the Bayesian NPIV-based estimates to be much closer to the previous estimates from the engineering literature, particularly for Eastbound SR-91. However, a substantial difference between our Bayesian NPIV estimate and the one reported in the engineering literature for Westbound SR-24 can be attributed to the bias in previous estimates due to minute-to-minute fluctuations in flow which might have caused due to the use of only a few days of observations \citep{Anderson2020}. We emphasise that our causal estimates of capacity are more representative of the actual capacity value as they are based on several months of observations and also adjusted for any potential confounding biases.

\subsubsection{Activation of the bottleneck and estimated capacity-drop}
\label{S:5.1.2}

Table \ref{tab:Comp} summarises the estimated capacity-drop in each highway section obtained via the non-IV-based and IV-based estimators. Table \ref{tab:Occ} reports the occupancy values corresponding to the drop in capacity.  

For Westbound SR-24, we observe a statistically significant drop of 27.42 percent in capacity at an occupancy level of 17.56 (the point of bottleneck activation) using the Bayesian NP estimator (see Figure \ref{fig:SR24W}(a)).  Figure \ref{fig:SR24W}(a) also shows a recovery in capacity to a value close to 600 vehicles per five-minutes following which there is a huge drop of about 50 percent. However, the evidence of recovery, followed by another drop, seems to be weak as the credible bands in this region are not tight. On the other hand, the Bayesian NPIV estimates show a statistically significant drop in flow from 512 to 473 vehicles per five minutes at an occupancy level of 17.27 (see Figure \ref{fig:SR24W}(b)). This fall in capacity corresponds to a statistically significant drop of 7.80 percent. 

For Eastbound SR-91, the results of the Bayesian NP-based estimation shown in Figure \ref{fig:SR91E}(a) suggests a statistically significant drop of 14.02 percent in capacity at an occupancy level of 16.13. However, the Bayesian NPIV-based estimate shown in Figure \ref{fig:SR91E}(b) illustrates that there is no statistically significant drop in capacity. From this figure, we note a lack of statistical evidence to suggest any change in flow beyond an occupancy level of 17.50.

For Eastbound SR-12, we observe a statistically significant drop of 11.54 percent in capacity at an occupancy level of 17.71 using the Bayesian NP-based estimator (see Figure \ref{fig:SR12E}(a)). This Bayesian NP estimate of capacity drop is close to the Bayesian NPIV-based estimate (10.92 percent), which occurs at an occupancy level of 17.08 (see Figure \ref{fig:SR12E}(b)).

The above comparison shows that the capacity-drop is overestimated by the Bayesian NP-based estimator in two out of the three sections, but the Bayesian NPIV and Bayesian NP estimates concur for the third section. In contrast to a recent study in the economics literature by \cite{Anderson2020} that rules out the existence of a statistically significant capacity-drop in highway bottlenecks, we do find sufficient statistical evidence of capacity-drop in two out of the three sections. Thus, the existence of capacity drop (or two-capacity phenomenon) must be evaluated on a case-by-case basis. It is worth noting that consistent with \cite{Cassidy1999a}, we also find that in all the three sections, the level of occupancy corresponding to a drop in capacity are almost similar, that is, $\sim17.0$. 

\subsubsection{The estimated congested (or hypercongested) regime of the flow-occupancy curve}
\label{S:5.1.3}

Figures \ref{fig:SR24W}(c), \ref{fig:SR91E}(c) and \ref{fig:SR12E}(c)) illustrate that the non-IV-based estimators (Bayesian NP and POLS) underestimate the congested regime of the flow-density curve that lies beyond the capacity point. These figures show that non-IV-based estimators exhibit a statistically significant backward bending relationship between flow and occupancy following the capacity-drop for all the three sections. However, the IV-based based estimator rules out the possibility of any statistically significant changes in flow with increase in occupancy following the capacity-drop in two out of the three sections. For Eastbound SR-12, we do find a statistically significant evidence of a backward bending relationship up to a certain level of occupancy after the initial capacity-drop.

It is worth noting that the IV-based estimates statistically reinforce some previous observations from the engineering literature. Our estimates support the study by \cite{Daganzo1999}, which presents empirical evidence to show that the entire backward bending part of the fundamental diagram for a highway section arises due to the presence of downstream disturbances or weather related events that might affect average driving behaviour. Such obstructions give rise to a predictable flow-density or flow-speed relationship; otherwise, the capacity of the section does not drop even when the demand is high \citep{Daganzo1999}. Our Bayesian NPIV estimates support this observation -- we do not observe a backward bending relationship for Westbound SR-24 and Eastbound SR-91 as these sections are perfectly isolated from any downstream bottlenecks, but a statistically significant backward bending relationship is evident for Eastbound SR-12 because SR-12 merges with I-80 just downstream of the analysed site. 

Our empirical IV-based estimates of the flow-occupancy curve are also consistent with the amendment in the fundamental speed-flow relationship proposed in the economics literature by \cite{Verhoef2001}. Based on car following theory, \cite{Verhoef2001} found that the entire backward bending part of the speed-flow relationship is dynamically unstable. Thus, consistent with \cite{Daganzo1999}, the amendment by \cite{Verhoef2001} suggests the absence of a backward bending part in the speed-flow curve, instead points towards a constant outflow from a highway section upon onset of congestion.

These results thus indicate the presence of large endogeneity biases in the non-IV-based estimates of the flow-occupancy curve (particularly in the congested regime), and thus, the advantages of adopting NPIV are apparent.

\subsection{Robustness Tests}
\label{S:5.2}

\subsubsection{Distribution of Errors}
\label{S:5.2.1}

\begin{figure}[H]
    \centering
        \begin{subfigure}{0.5\textwidth}
            \centering
            \includegraphics[width=1\textwidth]{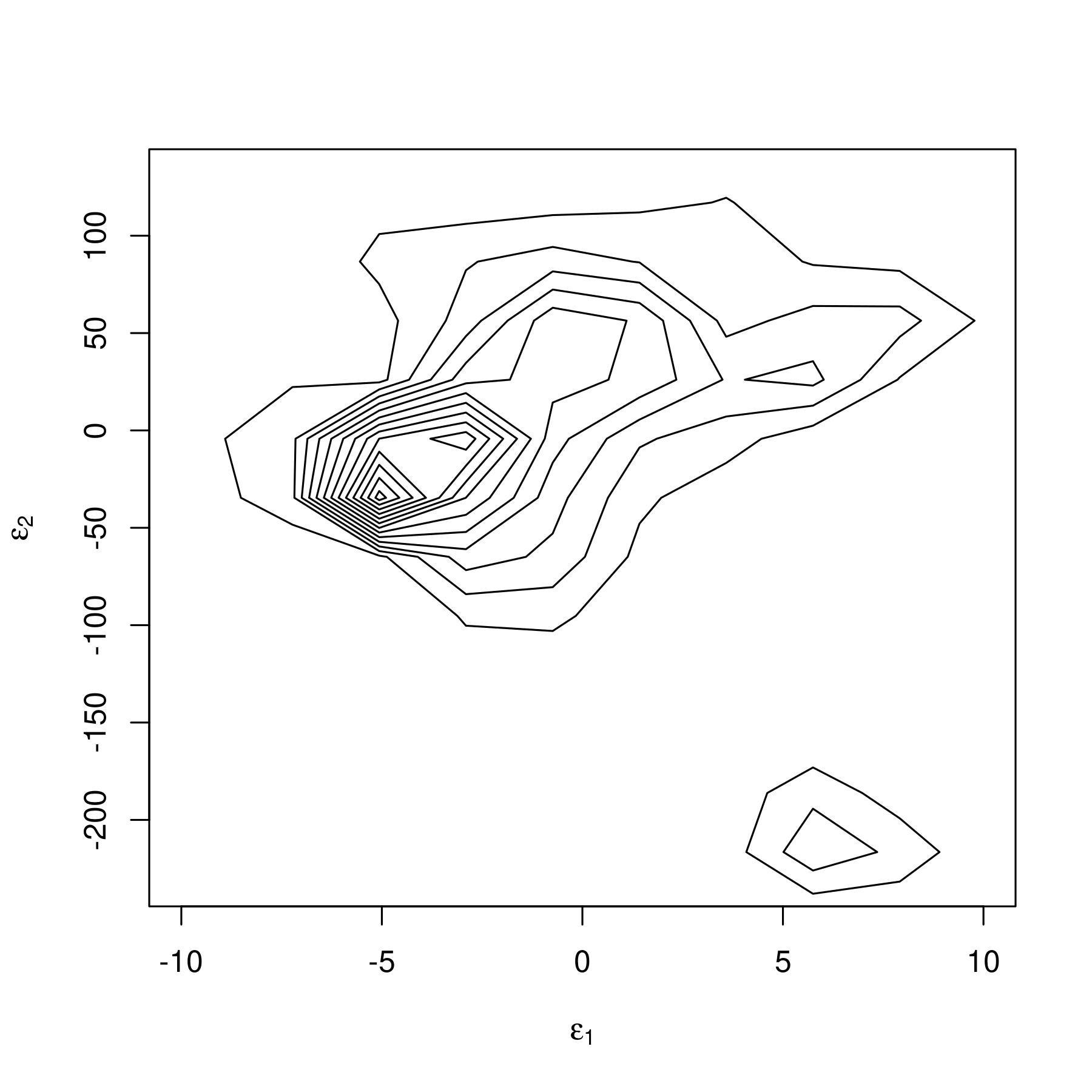}
            \caption{Westbound SR-24}
        \end{subfigure}%
        \hfill
        \begin{subfigure}{0.5\textwidth}
            \centering
            \includegraphics[width=1\textwidth]{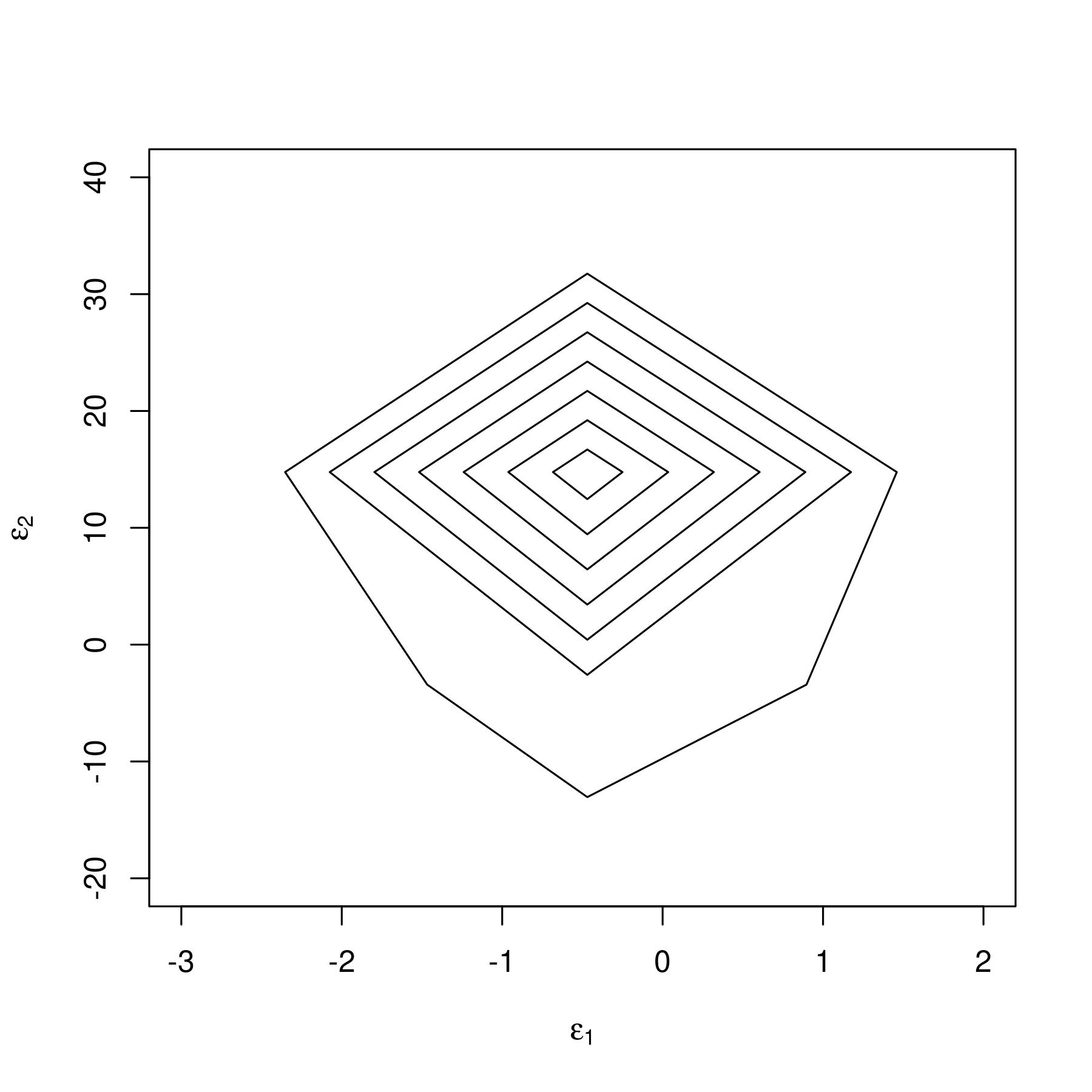}
            \caption{Eastbound SR-91}
        \end{subfigure}
        \hfill
        \begin{subfigure}{0.5\textwidth}
            \centering
            \includegraphics[width=1\textwidth]{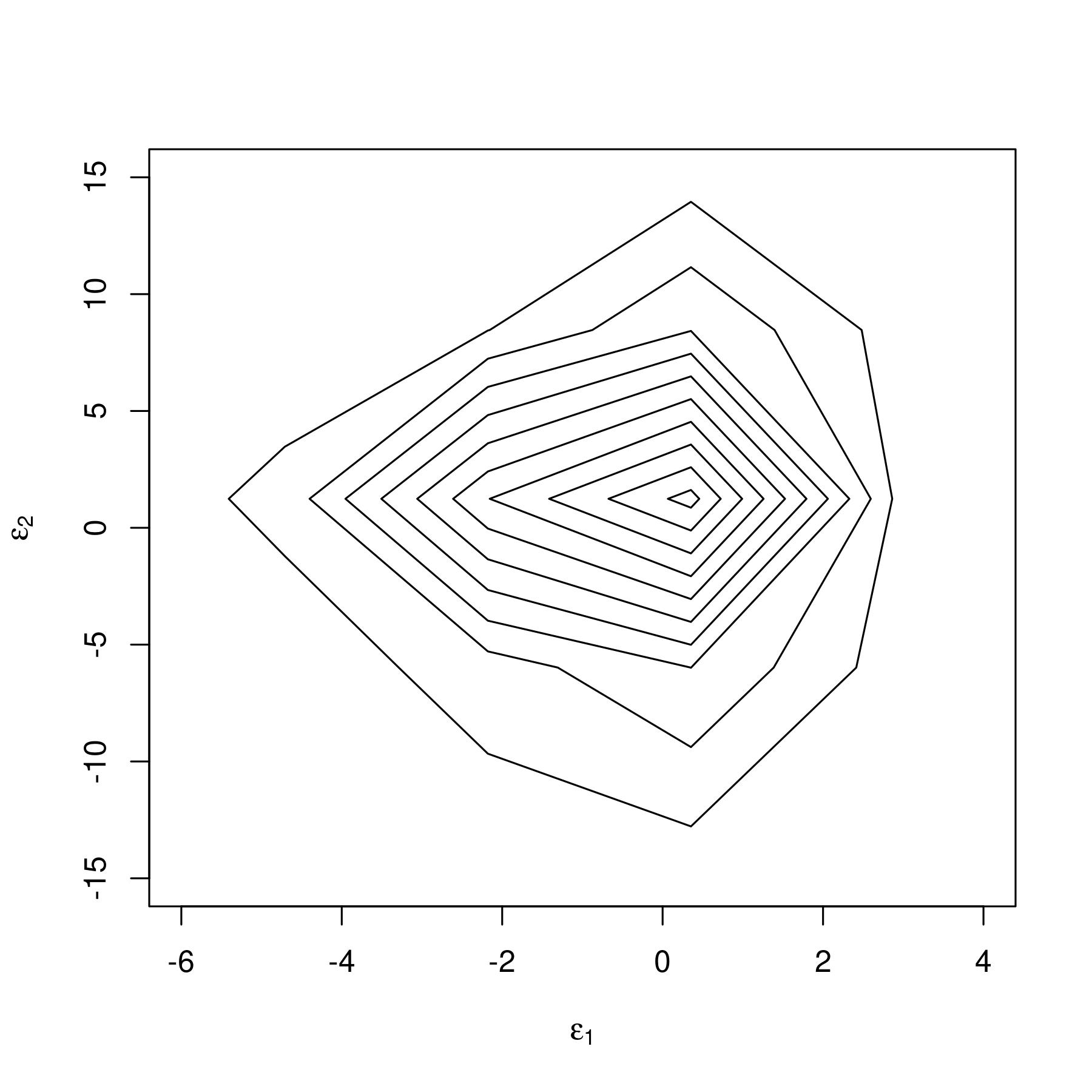}
            \caption{Eastbound SR-12}
        \end{subfigure}
    \caption{Distribution of errors.}
    \label{fig:Err_Dist}
\end{figure}

Figure \ref{fig:Err_Dist} shows the contour plot of the joint distribution of errors from the first stage ($\epsilon_1$) and the second stage ($\epsilon_2$). These figures show that the joint error distribution is either uni-modal asymmetric or bi-modal.

These results suggest that the estimates of $S(.)$ and inference could have poor statistical properties if the error is assumed to follow a uni-modal symmetric and thin-tailed Gaussian error distributions. The adopted Bayesian NPIV method addresses all these potential challenges by allowing for a flexible distribution of errors, instead of assuming a restrictive parametric error distribution.

\subsubsection{Relevance of Instruments}
\label{S:5.2.2}

\begin{figure}[tp]
    \centering
        \begin{subfigure}{0.5\textwidth}
            \centering
            \includegraphics[width=1\textwidth]{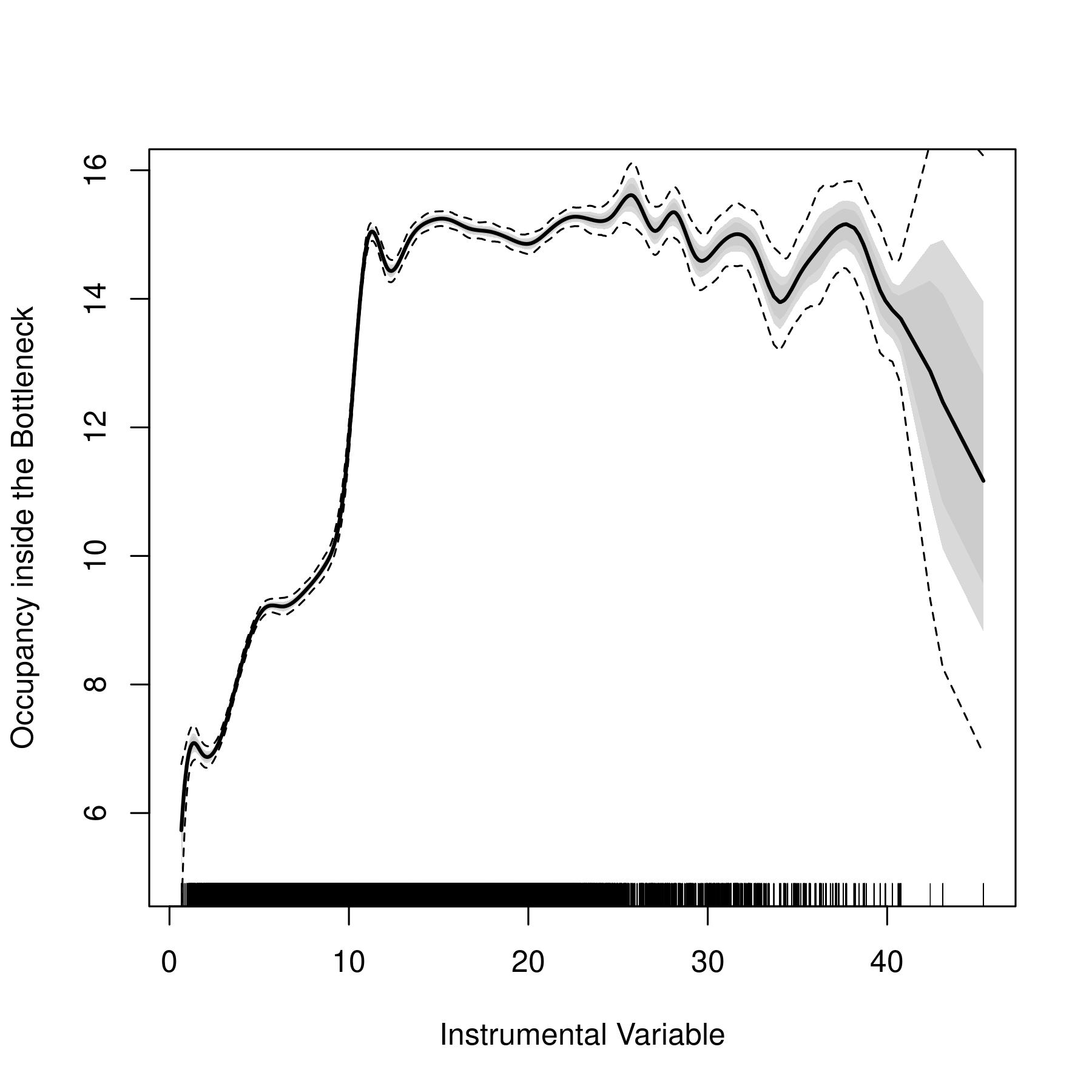}
            \caption{Westbound SR-24}
        \end{subfigure}%
        \hfill
        \begin{subfigure}{0.5\textwidth}
            \centering
            \includegraphics[width=1\textwidth]{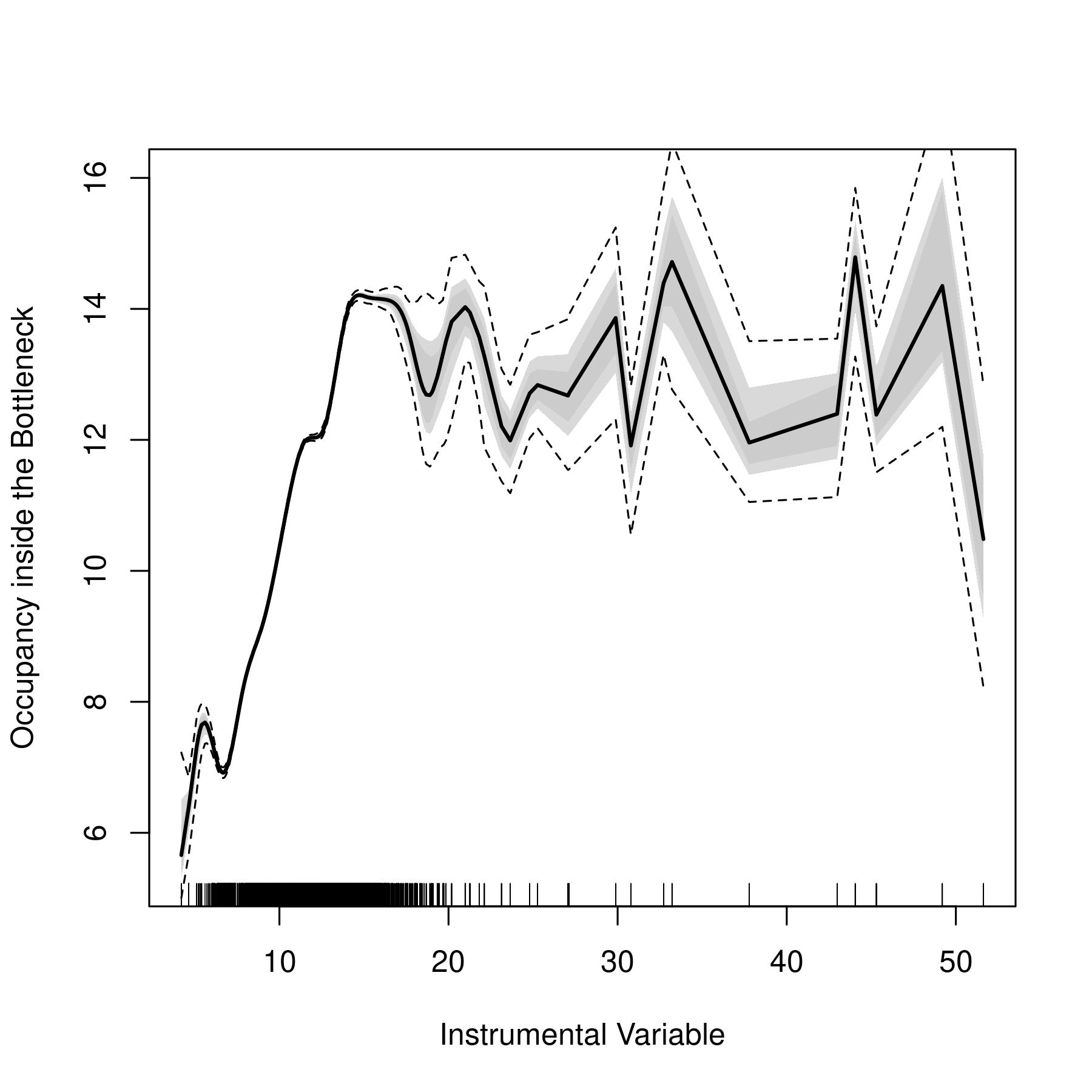}
            \caption{Eastbound SR-91}
        \end{subfigure}
        \hfill
        \begin{subfigure}{0.5\textwidth}
            \centering
            \includegraphics[width=1\textwidth]{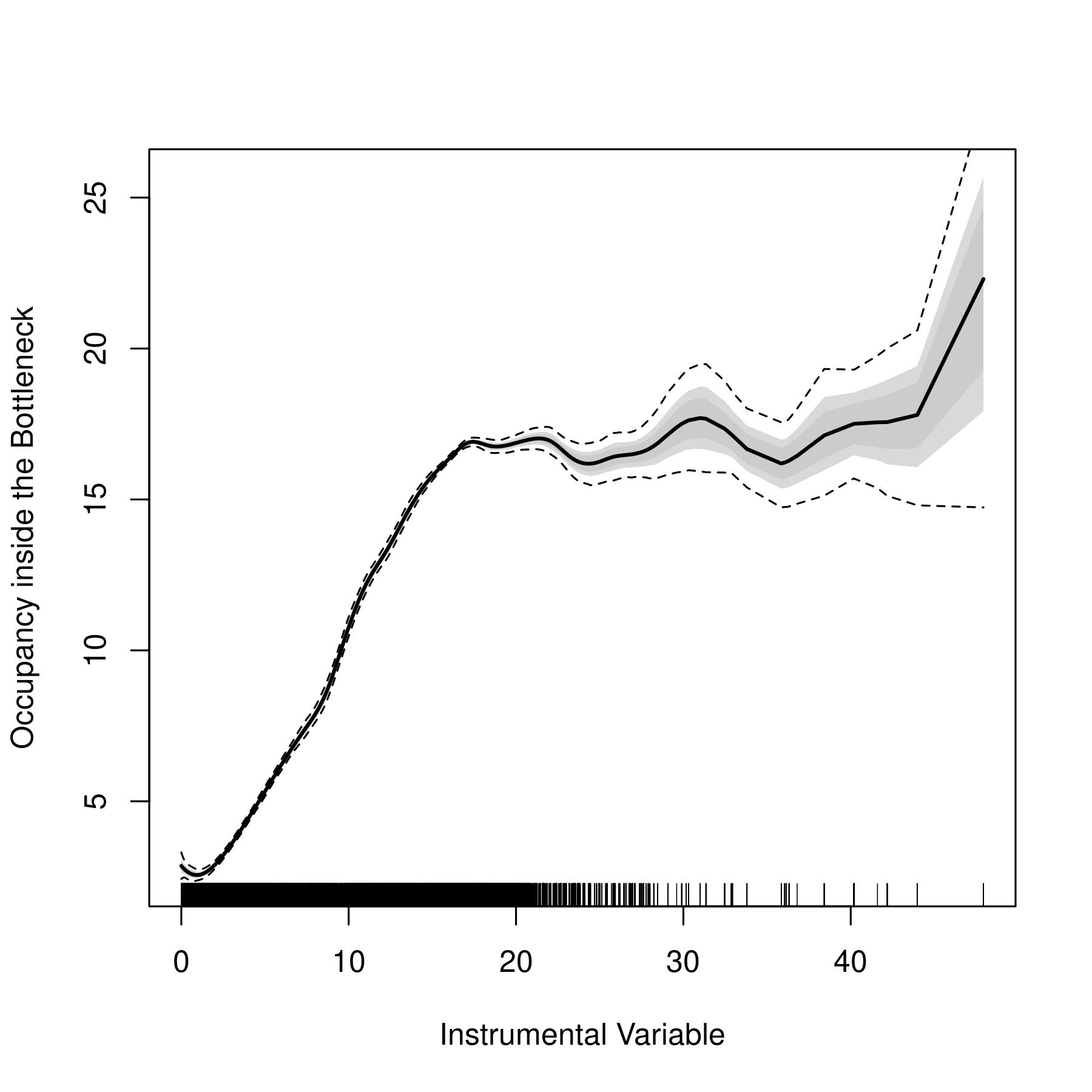}
            \caption{Eastbound SR-12}
        \end{subfigure}
    \caption{Relevance of instruments.}
    \label{fig:Inst_Str}
\end{figure}

Figure \ref{fig:Inst_Str} illustrates the results (that is, the estimated $h(.)$) from regression of the endogenous covariate on the instrument for the three highway sections.

These figures show a strong correlation between the instrument and the endogenous covariate for values of the IV less than 15, but  appears relatively weaker in the remaining domain of IV for SR-24 and SR-91. We thus carry out complimentary weak instrument tests to evaluate the relevance of the chosen instruments. 

To this end, we use the traditional F-tests \cite{Stock2005} at different parts of the IV's domain to test for the relevance of the chosen IV at a local level. Specifically, we divide the IV's domain into two bins and carry out the weak instrument test in each bin.  The corresponding F-statistics values, along with the one for the entire domain of the IV, are reported in Table \ref{tab:Ftest}. For all three study sites, F-statistics values across all considered domains of the IV are above the critical value of 10, as reported in \cite{Stock2005}. 

\begin{table}[H]
\caption{Summary of results from the Stock and Yogo instrument F-test.}
\label{tab:Ftest}
\centering
\begin{adjustbox}{width= 0.9\textwidth}
\begin{tabular}{L{4cm} C{4cm} C{3cm} C{3cm}} 
\hline
& \multicolumn{3}{c}{F-statistic} \\
Highway Section & for full support of IV & for IV $\leq 15$ & for IV $> 15$\\
\hline
Westbound SR-24 & $5.18 \times 10^4$ & $2.76 \times 10^4$ & 3179.00 \\ 
Eastbound SR-91 & $1.98 \times 10^4$ & $3.45 \times 10^4$ & 56.58 \\ 
Eastbound SR-12 & $4.82 \times 10^4$ & $3.75 \times 10^4$ & 182.10 \\ 
\hline
\end{tabular}
\end{adjustbox}
\end{table}

Thus, Figure \ref{fig:Inst_Str} and results from the Stock and Yogo weak instrument F-test summarised in Table \ref{tab:Ftest} provide supporting evidence that the selected IVs satisfy the relevance condition. 

\section{Conclusions and Future Work}
\label{S:6}

The contributions of this research are two-fold. Our methodological contributions reside in developing a comprehensive understanding of the fundamental relationship of traffic flow in a highway section by adopting a causal econometric framework to determine a novel causal relationship between traffic flow and occupancy in a highway section with a downstream bottleneck. We apply a Bayesian non-parametric instrumental variables (NPIV) estimator on data from three highway bottlenecks in California. The use of NPIV is attractive as it allows us to capture non-linearities in the fundamental  relationship with a fully flexible non-parametric specification and adjusts for confounding bias via the inclusion of relevant and exogenous instruments. Such confounding biases may occur because of many external observed or unobserved factors such as  driver behaviour, heterogeneous vehicles, weather and demand, that are correlated with both observed traffic variables. We thus deliver a more robust characterisation of traffic flow in a highway section that is reproducible and is not sensitive to these extraneous influences. As a by-product of the estimation, we produce novel quantitative estimates of capacity drop in the three bottlenecks.

Our theoretical contributions emerge from reconciling the economics and engineering approaches to estimate the empirical fundamental relationship of traffic flow. One prominent economic approach is based on a demand-supply framework where users of the highway section are treated as suppliers of travel in the section and outflow from the highway section in turn represents the travel supplied. However, we note that the equivalence of the fundamental relationship of traffic flow and the supply curve for travel in a highway section can only be considered under stationary state traffic conditions, which seldom exist particularly under congested traffic conditions. We thus argue that the demand-supply framework may lead to misrepresentation in developing a causal understanding of the empirical fundamental diagram. We instead adopt causal statistical modelling within the engineering framework which is based on the physical laws that govern the movement of vehicles in a traffic stream.

The above themes are important as a recent study in the economics literature examines the changes in outflow with increasing demand for three different highway bottlenecks in California and finds no evidence of drop in capacity or in other words, hypercongestion during periods of high demand. The study concludes that the fundamental (flow-density or flow-speed) diagram for a highway section should not exhibit a backward bending part and also questions the applicability of traffic control measures and congestion pricing policies that are aimed at regulating demand to avoid hypercongestion. Based on our estimated causal fundamental relationship, we re-evaluate the existence of capacity-drop in highway bottlenecks, which is a well-established phenomenon in the engineering literature.

Our empirical results show a statistically significant decrease in flow upon activation of the bottleneck in two out of three analysed bottlenecks, thus supporting the existence of capacity drop. The estimated capacity-drop varies on a case-to-case basis depending upon the geometry of the bottleneck as well as the characteristics of the average traffic stream passing through it. However, after this drop in capacity, we do not find sufficient statistical evidence to support any changes in flow with further increase in occupancy in isolated highway sections. We thus argue that as the flow through the bottleneck remains constant following the capacity-drop, the flow-occupancy curve is not actually backward bending. However, a statistically-significant backward bending relationship exists only when the highway section is not perfectly isolated from downstream obstacles that cause traffic flow through the section to decrease over occupancy in a predictable way.

It is important to note that the empirical results discussed in this paper apply only to a highway section with a standard bottleneck. These results are encouraging and the framework can be directly adopted to estimate a causal model of traffic flow for a uniform highway section. Our theoretical conclusions on the association between the fundamental relationship and the analysis of travel supply applies to both of these scenarios, that is, highway section with or without a bottleneck.

Our causal estimates of the fundamental relationship are crucial from a policy point of view in case of the design of highways and devising traffic control strategies, as these estimates provide a more generalised and robust characterisation of the traffic flow in a highway section and adjusts for any potential confounding biases. Our causal models are, therefore, more suited for standard reference manuals like the highway capacity manual (HCM) and the UK-CoBA. Our theoretical and empirical conclusions also have important implications for deriving highway tolls and congestion pricing policies, which we plan to undertake in future work.

\vspace{2cm}

\noindent \textbf{Declarations of interest}

\noindent None.

\bigskip

\noindent \textbf{Funding}

\noindent This research did not receive any specific grant from funding agencies in the public, commercial, or not-for-profit sectors.

\bigskip

\noindent \textbf{Acknowledgement}

\noindent Prateek Bansal is supported by the Leverhulme Trust Early Career Fellowship. Authors are also thankful to Manuel Wiesenfarth for helping with queries related to the Bayesian non-parametrics and implantation.

\newpage

\bibliographystyle{agsm}
\bibliography{Hypercongestion}

\newpage

\section*{Appendix A}

In this section, we demonstrate the potential sources of confounding discussed in Section \ref{S:4.2} in the fundamental relationship of traffic flow in mathematical terms.

\subsection*{Omitted Variable Bias}

To illustrate the endogeneity bias due to omitted covariates, we simplify equation \ref{eq:spec}, where we consider that $S(.)$ has a linear specification, that is $S(\mathbf{o})=\mathbf{o}\beta$. We suppose that $\mathbf{\delta}=\mathbf{w}\alpha$, where $\mathbf{w}$ represents, say, driving and vehicular characteristics. For the notational simplicity, we drop time-day subscripts and superscripts. We, thus, have a data generating process given by:

\begin{equation}
\label{eq:dgp}
\mathbf{q} = \mathbf{o}\beta + \mathbf{w}\alpha + \mathbf{\xi},
\end{equation}

\noindent where $\mathbf{q}$ is an $N \times 1$ vector of dependent variables, $\mathbf{o}$ and $\mathbf{w}$ are $N \times 1$ and $N\times K$ matrices and $\mathbf{\xi}$ is an $N \times 1$ error vector that is assumed to be uncorrelated with $\mathbf{o}$ and $\mathbf{w}$. Application of a standard regression technique such as an ordinary least squares (OLS) estimation of $\mathbf{q}$ on $\mathbf{o}$ and $\mathbf{w}$ yields consistent parameter estimates of $\alpha$ and $\beta$ \footnote{Note that an estimator $\hat{\beta}$ is said to be consistent for $\beta$ if it converges in probability to the true value $\beta$, that is, plim$(\hat{\beta}) \rightarrow \beta$.}.

Suppose instead that $\mathbf{w}$ is omitted from the equation and $\mathbf{q}$ is regressed on $\mathbf{o}$ alone. Then $\mathbf{w}\alpha$ becomes a part of the error term and the estimated model becomes:

\begin{equation*}
\label{eq:dgp_o}
\mathbf{q} = \mathbf{o}\beta + (\mathbf{w}\alpha + \mathbf{\xi}),
\end{equation*}

\noindent where $\mathbf{w}\alpha + \mathbf{\xi}$ is the new error term. The OLS estimator of $\beta$ equals:

\begin{equation*}
\label{eq:beta_ols}
\begin{split}
\beta_{OLS} &= \mathbf{(o'o)}^{-1}\mathbf{o'q} \\
&= \mathbf{(o'o)}^{-1}\mathbf{o'(o\beta + w\alpha + \xi)} \\
&= \mathbf{(o'o)}^{-1}\mathbf{o'o\beta} + \mathbf{(o'o)}^{-1}\mathbf{o'w\alpha} + \mathbf{(o'o)}^{-1}\mathbf{o'\xi} \\
&= \beta + (N^{-1}\mathbf{o'o}^{-1})(N^{-1}\mathbf{o'w})\alpha + (N^{-1}\mathbf{o'o})^{-1}(N^{-1}\mathbf{o'\xi}) \\
\end{split}
\end{equation*}

Under the assumption that $\mathbf{o}$ is uncorrelated with $\mathbf{\xi}$, the final term has probability limit zero. However, because, $\mathbf{o}$ is correlated with $\mathbf{w}$,

\begin{equation*}
\label{eq:beta_ols_plim}
\textrm{plim} [\beta_{OLS}] = \beta + \delta \alpha
\end{equation*}

\noindent where, $\delta = \textrm{plim}[(N^{-1}\mathbf{o'o}^{-1})(N^{-1}\mathbf{o'w})]$ is the probability limit of the OLS estimator in the regression of the omitted regressor ($\mathbf{w}$) on the included regressors ($\mathbf{o}$). This inconsistency is called omitted variable bias, which exists as long as the omitted regressor is correlated with the included regressors. In general the inconsistency could be positive or negative. A positive bias exists if the correlation between $\mathbf{o}$ and $\mathbf{w}$, that is, $\delta$ and that between $\mathbf{q}$ and $\mathbf{w}$, that is, $\alpha$ are both either positive or negative, that is, $\alpha \delta > 0$. If these correlations are of opposite sign, that is, $\alpha \delta < 0$, the bias is negative. For instance, if $\mathbf{w}$ represents the risk-taking ability of drivers, we may expect a positive correlation between $\mathbf{o}$ and $\mathbf{w}$ as well as $\mathbf{q}$ and $\mathbf{w}$, resulting in positive bias due to omission of drivers' risk taking abilities. This is because we may expect an average population of risk taking drivers to drive at smaller headways or higher densities even at very high speeds, thus resulting into larger flows. 

In Table \ref{tab:confounders}, we enlist various confounders for the fundamental relationship based on the literature (refer Section \ref{S:2.1}) and their expected correlations with occupancy and flow.

\begin{table}[H]
\caption{Various sources of confounding in the fundamental relationship.}
\label{tab:confounders}
\centering
\begin{adjustbox}{width= 0.9\textwidth}
\begin{tabular}{L{7cm} C{4cm} C{4cm}} 
\hline
Confounder & Expected correlation with flow & Expected correlation with occupancy \\
\hline
Risk-taking behaviour of drivers & + & + \\ 
Risk-averse behaviour of drivers & - & - \\ 
Vehicle accelerations & - & + \\ 
Vehicle decelerations & + & - \\
Lane change manoeuvres & +/- & +/- \\ 
Vehicle lengths & +/- & +/- \\ 
Detector-level (measurement) errors & +/- & +/- \\
Weather conditions & +/- & +/- \\
Other characteristics of demand & +/- & +/- \\
\hline
\end{tabular}
\end{adjustbox}
\end{table}

\subsection*{Reverse Causality}

To illustrate bias due to reverse causality, we further simplify the data generating process in equation \ref{eq:dgp} as follows:

\begin{equation}
\label{eq:dgp_new}
\mathbf{q} = \mathbf{o}\beta + \mathbf{\xi},
\end{equation}

To obtain an unbiased estimate of $\beta$ via OLS, the Gauss Markov condition of zero conditional mean of errors, that is, $E[\xi|o] = 0$, or in other terms, $\textrm{Cov}[\xi,o] = 0$,must be satisfied. In case of reverse causality, there exists another data generating process given by:

\begin{equation}
\label{eq:dgp_new_rev}
\mathbf{o} = \mathbf{q}\gamma + \mathbf{\psi},
\end{equation}

Consequently, we have,

\begin{equation*}
\label{eq:cov_rc}
\begin{split}
\textrm{Cov}[\xi,o] &= \textrm{Cov}[\xi,(q\gamma + \mathbf{\psi})] \\
&= \gamma\textrm{Cov}[\xi,q] \quad \textrm{assuming that} \quad \xi \perp \psi \\
&= \gamma\textrm{Cov}[\xi,(o\beta+\xi)] \\
&= \gamma\textrm{Cov}[\xi,o\beta]+\textrm{Var}(\xi) \\
&\neq 0 \\
\end{split}
\end{equation*}

Thus, the zero conditional mean assumption of errors is violated and OLS may result into a biased estimate of $\beta$.

\end{document}